\documentclass[preprint]{aastex62}

\hypersetup{linkcolor=red,citecolor=cyan,filecolor=blue,urlcolor=magenta}
\usepackage{subfigure}
\usepackage{color}
\shorttitle{GB 1508+5714: the {\it first} identified $\gamma$-ray blazar beyond redshift 4}
\shortauthors{Liao et al.}

\begin{document}

\title{Detections of simultaneous brightening of $\gamma$-ray and optical emissions of a distant blazar GB~1508+5714 at redshift 4.3}

\correspondingauthor{Neng-Hui Liao,}
\email{nhliao@gzu.edu.cn}

\author[0000-0001-6614-3344]{Neng-Hui Liao}
\affiliation{Department of Physics and Astronomy, College of Physics, Guizhou University, Guiyang 550025, China}

\author{Shang Li}
\affiliation{Key Laboratory of Dark Matter and Space Astronomy, Purple Mountain Observatory, Chinese Academy of Sciences, Nanjing 210034, China}

\author{Zhen-Feng Sheng}
\affiliation{Key laboratory for Research in Galaxies and Cosmology, Department of Astronomy, The University of Science and Technology of China, Chinese Academy of Sciences, Hefei, Anhui 230026, China}

\author{Yi-Zhong Fan}
\affiliation{Key Laboratory of Dark Matter and Space Astronomy, Purple Mountain Observatory, Chinese Academy of Sciences, Nanjing 210034, China}

\begin{abstract}
GB 1508+5714 is a high-redshift blazar ($z$ = 4.3) and a spectrally soft $\gamma$-ray source has been detected in its direction. By analyzing 11.4-yr {\it Fermi}-LAT data, significant long-term variability of the $\gamma$-ray source is confirmed. More importantly, a $\gamma$-ray emission flare appeared in an epoch of several tens of days in year 2018, when the flux is about four times of the value from the global fit. Meanwhile, optical flares displayed in both $r$- and $i$-bands from the {\it Zwicky} Transient Facility light curves. Detections of the simultaneous $\gamma$-ray and optical brightening provide a decisive evidence to pin down the association between the $\gamma$-ray source and GB 1508+5714, which makes it the {\it first} identified $\gamma$-ray blazar beyond redshift 4. A broadband spectral energy distribution in the high flux state is constructed and the origin of the multiwavelength brightening is also briefly discussed. Upcoming wide-deep-fast optical telescopes together with the $\gamma$-ray sky surveyors will shed lights on the role that the AGN jets play in the early cosmic time.
\end{abstract}

\keywords{galaxies: active -- galaxies: high-redshift -- galaxies: jets -- gamma rays: galaxies -- quasars: individual (GB 1508+5714)}

\section{Introduction} \label{sec:intro}
Benefited from the strongly boosted jet emission \citep{1978bllo.conf..328B,2019ARA&A..57..467B}, blazars are bright beacons \citep[e.g.,][]{2014Natur.515..376G} and capable to be detected even in the early cosmic time \citep[e.g.,][]{2004ApJ...610L...9R,2006AJ....132.1959R}. The broadband jet emission is characterized by a universal two-bump spectral energy distribution (SED) structure in log$\nu$F$\nu$-log$\nu$ plot, where one is widely accepted as synchrotron emission while the other one extending to the $\gamma$-ray domain. In the leptonic scenarios, $\gamma$-ray emissions of blazars are usually explained as inverse Compton (IC) scattering of soft photons from either inside (the synchrotron self-Compton, or SSC, \citealt{1992ApJ...397L...5M}) and/or outside (external Compton, or EC, \citealt{1993ApJ...416..458D,1994ApJ...421..153S,2000ApJ...545..107B}) of the jet by the same population of relativistic electrons that are responsible for the synchrotron emission. In addition, the coincidence between the incoming of a sub-PeV neutrino event and multiwavelength flares in TXS 0506+056 suggests that the hadronic processes should be taken into account at least in some cases \citep{2018Sci...361.1378I}. High-redshift blazars are valuable targets for understanding the formation and growth of the first generation of super massive black holes as well as the cosmic evolution of AGN jets \citep{2010MNRAS.405..387G,2010A&ARv..18..279V}. Meanwhile, their emissions carry crucial information of the early universe. Particularly, the $\gamma$-ray emission of high-redshift blazars is valuable for probing the extragalactic background light (EBL, e.g., \citealt{2018Sci...362.1031F}). However, as a result of the faintness due to their large distances, the number of detected high-redshift (i.e. $z >$ 4) blazars (candidates) is limited \citep{2009A&A...495..691M}. In fact, it becomes rather challenging in the $\gamma$-ray domain. In the fourth $Fermi$ Large Area Telescope (LAT, \citealt{2009ApJ...697.1071A}) catalog (4FGL, \citealt{2020ApJS..247...33A}), all blazars beyond redshift 3 are flat-spectrum radio quasars (FSRQs). Since the peak of their high energy SED bump is beneath the lower energy threshold of {\it Fermi}-LAT, the decline of $\gamma$-ray emissions caused by the significant cosmic redshift is a major obstacle for $\gamma$-ray blazar detection at high redshifts.

Blazars are characterized by violent multiwavelength variability \citep[e.g.,][]{1997ARA&A..35..445U,2016ARA&A..54..725M}. Coincidences between their $\gamma$-ray flares and ones in other windows of electromagnetic radiation have been frequently detected \citep[e.g.,][]{2010Natur.463..919A,2014ApJ...783...83L}. In the perspective of the relatively limited angular resolution of the $\gamma$-ray observation, catching these correlated variations provide a decisive proof to support the association between the $\gamma$-ray source and its low-energy counterpart \citep[e.g.,][]{2011A&A...532A.150S,2016ApJS..226...17L}. In the high-redshift regime, based on the strong quasi-simultaneous IR and $\gamma$-ray flares, CGRaBS J0733+0456 ($z$ = 3.01) is proved as an identified $\gamma$-ray source \citep{2019ApJ...879L...9L}.  GB 1508+5714 was initially detected as a radio source at 5~GHz \citep{1992MNRAS.254..655P} and then identified as a high-redshift quasar ($z$ = 4.3, \citealt{1995MNRAS.273L..63H}).  Considering the flat radio spectrum and the high radio loudness \citep{1989AJ.....98.1195K} value, as well as the hard X-ray spectrum \citep{1995AJ....110.1551M,1997ApJ...484L..95M}, it has been suggested to be a blazar \citep{2009A&A...495..691M}. More importantly, a faint and spectrally soft $\gamma$-ray source (categorized as 4FGL J1510.1+5702 in 4FGL, \citealt{2020ApJS..247...33A}) cospatial with GB 1508+5714 is found and hence it is claimed as the most distant $\gamma$-ray blazar so far \citep{2017ApJ...837L...5A}.  By analyzing nearly the first 9 years $Fermi$-LAT data, the $\gamma$-ray source towards GB 1508+5714 is suggested to be likely variable \citep{2018ApJ...853..159L}. Meanwhile, its overall broadband SED is similar with those of $\gamma$-ray FSRQs, further supporting its blazar nature \citep{2020ApJ...889..164M}. However, due to the faintness and the soft $\gamma$-ray spectrum, and considering that the angular resolution of $Fermi$-LAT for sub-GeV photons is much worse than for GeV $\gamma$ rays\footnote{http://www.slac.stanford.edu/exp/glast/groups/canda\/lat\_Performance.htm}, its nature of $\gamma$-ray blazar can not be set in stone until detections of multiwavelength correlated variations.

In this Letter, we analyze the 11.4-yr of $Fermi$-LAT data as well as the Intermediate Palomar Transient Factory (iPTF) and {\it Zwicky} Transient Facility (ZTF) light curve data of GB 1508+5714, and investigate its multiwavelength variability properties (Section \ref{sec:data}), along with some discussions (Section \ref{sec:discu}). Here we take a $\Lambda$CDM cosmology with $H_{0}=67~{\rm km~ s^{-1}~Mpc^{-1}}$, $\Omega_{\rm m}=0.32$, and $\Omega_{\Lambda}=0.68$ \citep{2014A&A...571A..16P}.

\section{Data Analysis and Results} \label{sec:data}
\subsection{{\it Fermi}-LAT data}
Here the $Fermi$-LAT Pass 8 {\tt SOURCE} data (from MJD 54682 to MJD 58852) with energy range between 100 MeV and 500 GeV are collected and analyzed by the {\tt Fermitools} software version 1.2.23. The entire data set is filtered with the zenith angle cut ($< 90^{\circ}$) as well as the recommended quality-filter cuts ({\tt DATA\_QUAL==1 \&\& LAT\_CONFIG==1}). {\tt Unbinned} likelihood analyses implemented in the {\tt gtlike} task is used to extract the $\gamma$-ray flux and spectrum. The initial background model includes all 4FGL sources within 15$^{\circ}$ around 4FGL J1510.1+5702, together with the diffuse $\gamma$-ray emission templates (i.e. {\tt gll\_iem\_v07.fits} and {\tt iso\_P8R3\_SOURCE\_V2\_v1.txt}). During the likelihood analyses, parameters of all 4FGL sources lying within a $10^{\circ}$ region of interest centered at the location of 4FGL J1510.1+5702, as well as the normalizations of the two diffuse emission backgrounds are set free. The significance of a $\gamma$-ray source is quantified by the test statistic (TS, \citealt{1996ApJ...461..396M}), which is defined as $TS=-2{\rm ln}({L_{0}/L}$) where ${L}$ and $L_{0}$ are the maximum likelihood values for the model with and without target source, respectively. In the temporal analysis, the spectral parameters of background sources are frozen with the values from the global fit, unless they are bright or close to the target. Meanwhile, the faint (TS $<$ 10) background sources are removed from the model and then the likelihood analyses are performed again. If TS value of the target is lower than 10, the 95\% confidential level (C.L.) upper limit is calculated by the {\tt pyLikelihood  UpperLimits} tool instead of estimating the flux. 

\begin{figure}
\centering
\includegraphics[scale=0.35]{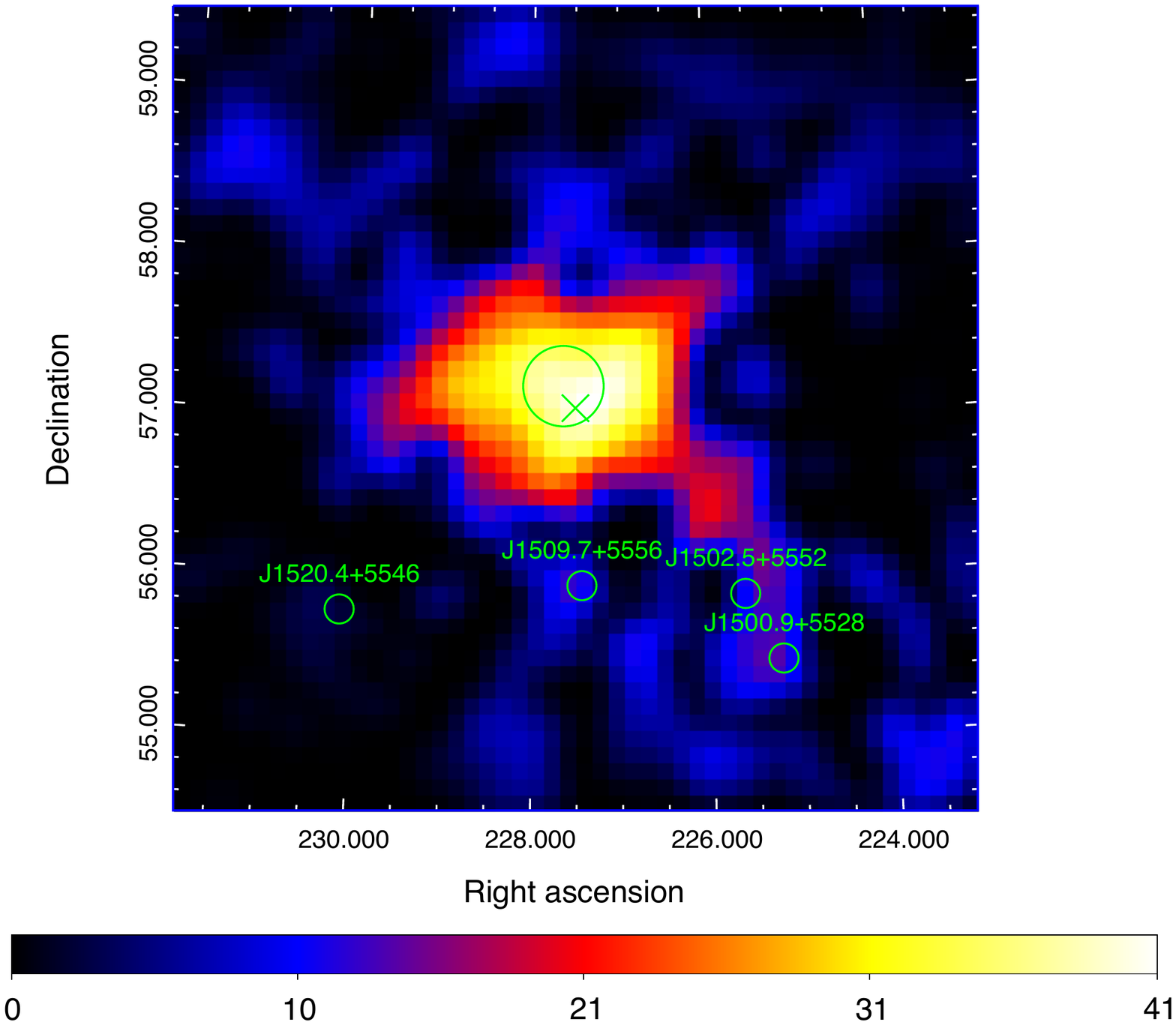}
\includegraphics[scale=0.35]{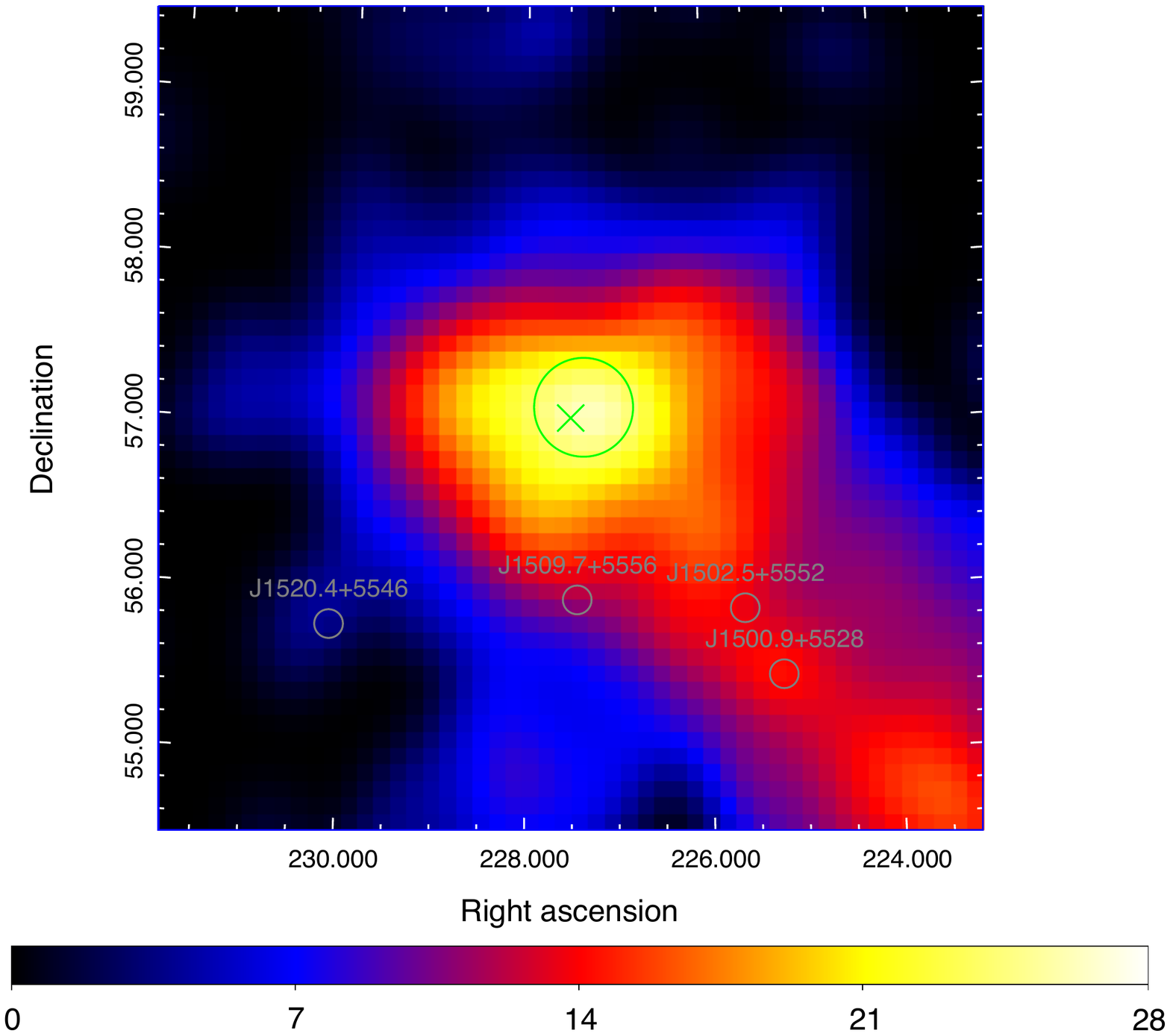}
\caption{$\gamma$-ray Residual (i.e. 4FGL J1510.1+5702 is not included in the analysis model file) TS maps for GB 1508+5714. {\it Left} panel: data time range between MJD 57602 and MJD MJD 58852 (i.e. the last 3.4-yr); {\it Right} panel: data time range between MJD 58217 and MJD 58308 (i.e. the 40th bin of the 3-month light curve). The scale of the TS maps is 10$\degr \times$10$\degr$ with 0.1$\degr$ per pixel. The maps are centered at 4FGL J1510.1+5702 and derived by using {\it Fermi}-LAT data between 0.1 and 500~GeV. The green X-shaped symbol represents the radio position of GB 1508+5714. The green circles are the 95\% C. L. error radii of the locations of the $\gamma$-ray source. Locations of the nearby background sources are also marked, along with their 4FGL names. Note that their TS values are lower than 10 in the 3-month analysis, and hence they are removed from the model file, colored as grey. }
\label{Fig.tsmap}
\end{figure}

The analysis of the entire 11.4-yr $Fermi$-LAT data yields a significant $\gamma$-ray source in the direction of GB 1508+5714. The $\gamma$-ray source has a rather soft spectrum, $dN/dE \propto E^{-(3.05 \pm 0.15)}$, consistent with the result from 4FGL \citep{2020ApJS..247...33A}. The radio location of GB 1508+5714 remains to be within the $\gamma$-ray 95\% C.L. error radius. Moreover, the TS value (TS = 78, 8.1$\sigma$) is doubled compared with it from 4FGL (TS = 37, 5.2$\sigma$, \citealt{2020ApJS..247...33A}\footnote{Note in 4FGL the summed-binned-likelihood analysis thread is adopted and the energy range of the selected $Fermi$-LAT data there is between 50~MeV and 100~GeV.}). The largely enhanced TS value suggests that a rise of the $\gamma$-ray emission likely appears after the first 8-year $Fermi$-LAT operation. Therefore, an individual analysis focusing on the last 3.4-yr $Fermi$-LAT data has been performed, and we find a robust $\gamma$-ray source (TS = 41, 5.6$\sigma$), as shown in Figure \ref{Fig.tsmap}. Its spectral index is constrained as 2.94 $\pm$ 0.18. Localization analysis gives an optimum location of R.A. 227.65$\degr$ and DEC. 57.182$\degr$ with a 95\% C. L. error radius of 0.25$\degr$, which overlaps with the radio location of GB 1508+5714. And there is no other blazars (candidates) \citep[e.g.,][]{2008ApJS..175...97H,2009A&A...495..691M} in such a region. We have checked whether there are new $\gamma$-ray sources (i.e. not included in 4FGL) close to 4FGL J1510.1+5702 emerging recently. In a short period, from MJD 58308 to MJD 58398, a new $\gamma$-ray source with TS value of 34 just 0.5$\degr$ away from the radio position of GB 1508+5714 appeared. Its optimum location is R.A. 226.755$\degr$ and DEC. 57.336$\degr$, with a 95\% C. L. error radius of 0.3$\degr$, which might associate with a radio source NVSS J150754+571723. After extracting this source, no significant $\gamma$-ray residual is found towards GB 1508+5714 then. In consideration of the proximity, {\it Fermi}-LAT data during such a 3-month epoch are eliminated during the entire and the last 3.4-yr $Fermi$-LAT data analyses.

\begin{figure}
\centering
\includegraphics[scale=0.5]{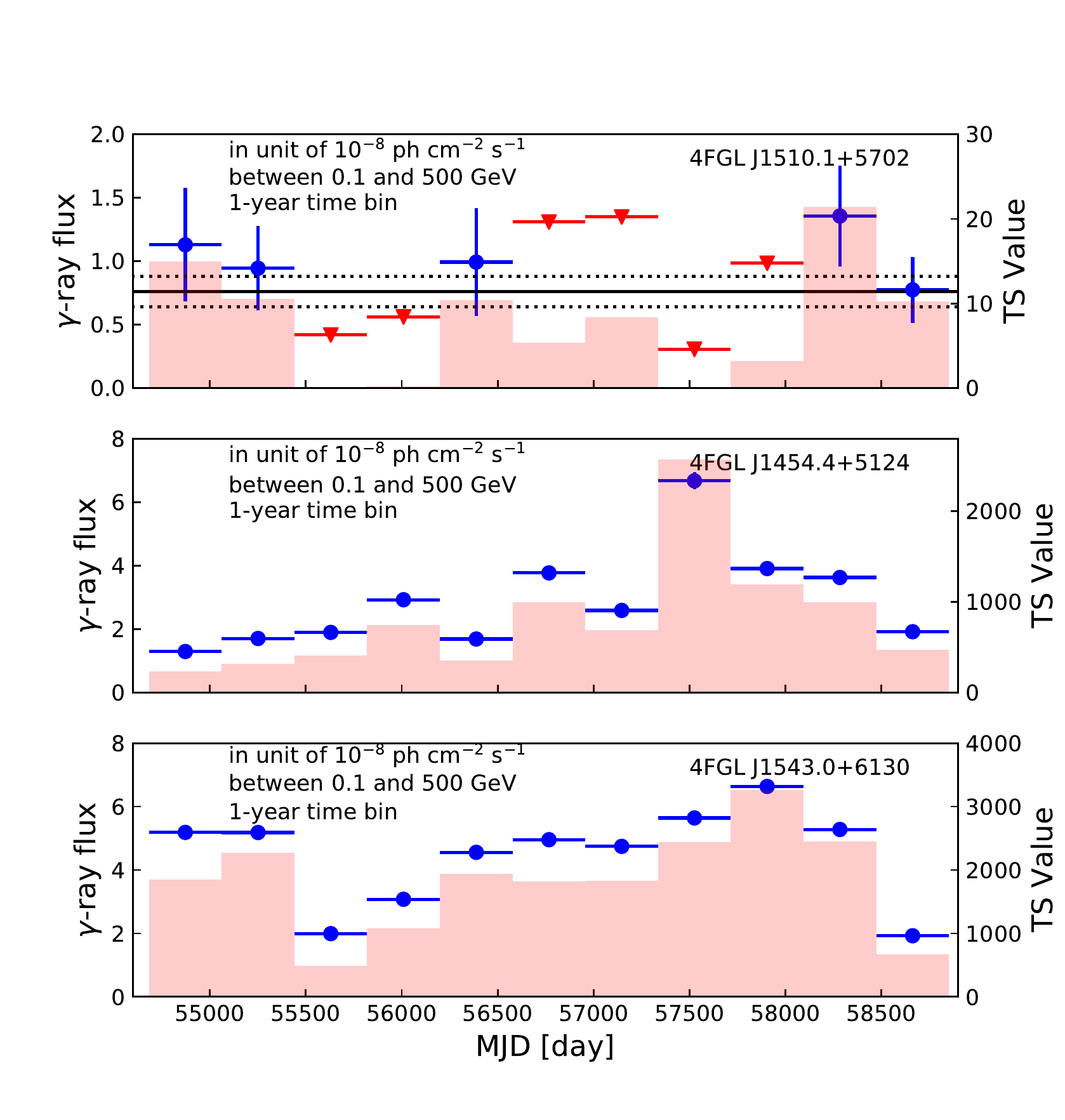}
\includegraphics[scale=0.5]{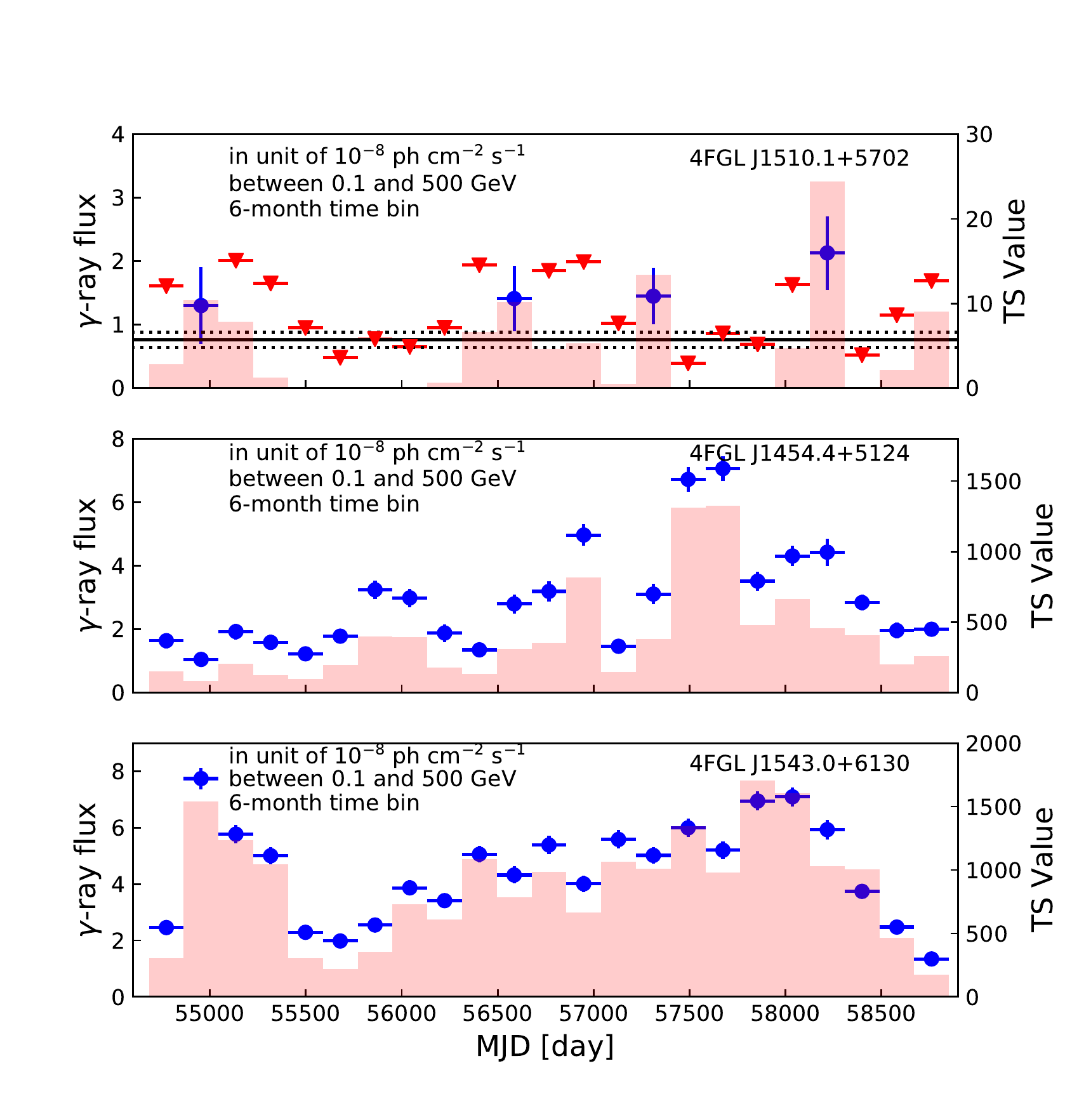}\\
\vspace{.01in}
\includegraphics[scale=0.5]{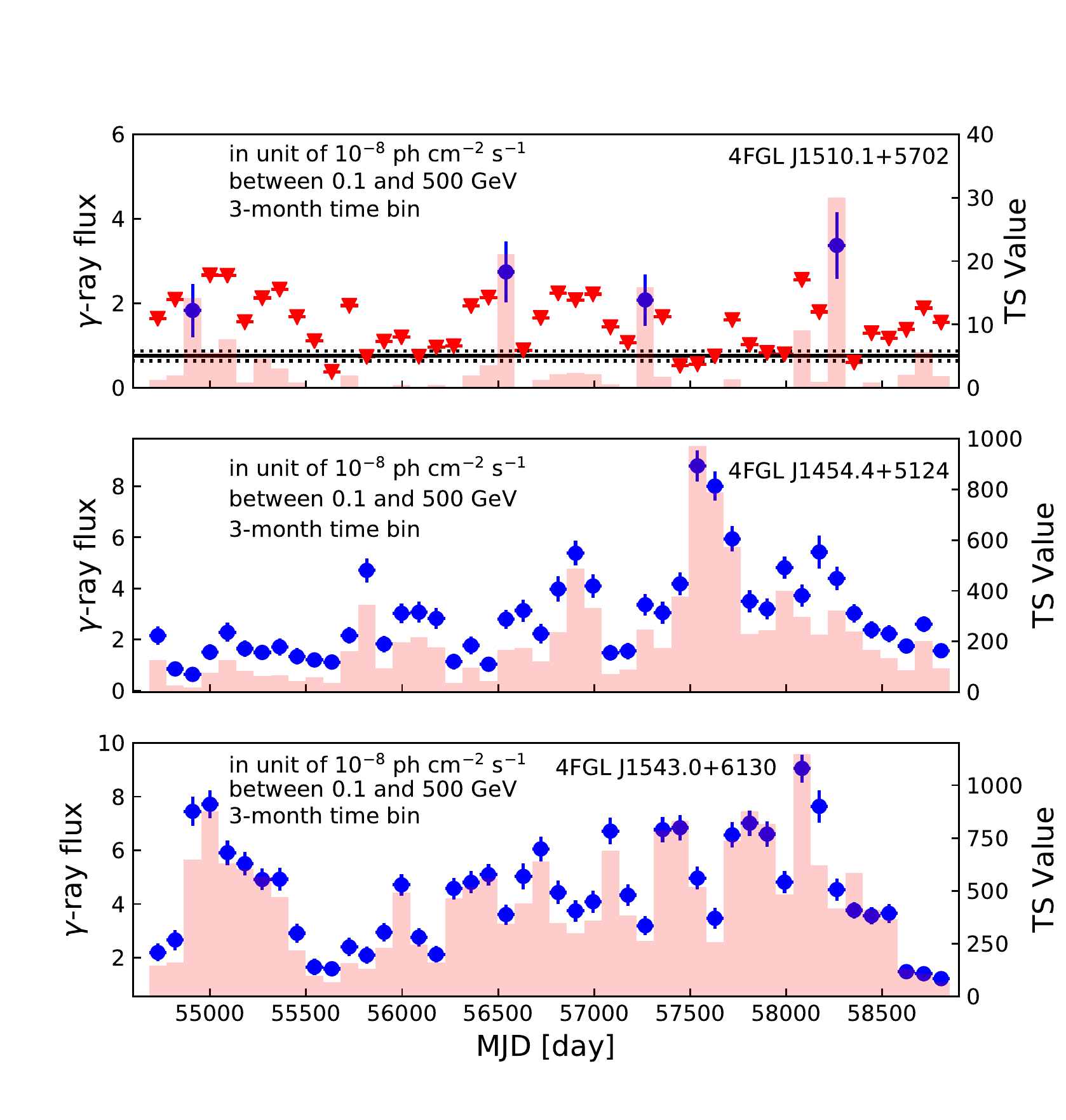}
\caption{$\gamma$-ray light curves of GB 1508+5714 as well as its neighbors 4FGL~J1454.4+5124 and 4FGL~J1543.0+6130. Blue points represent the $\gamma$-ray fluxes, while the red triangles are upper limits. Red bars are the corresponding TS values. The 11.4-year averaged $\gamma$-ray flux (solid line) and its 1$\sigma$ uncertainty (dotted lines) are also marked.}
\label{Fig.glc}
\end{figure}

Since the $\gamma$-ray source is relatively faint, firstly, a 1-year time bin light curve is extracted, see Figure \ref{Fig.glc}. Though in this case the $\gamma$-ray source is not well distinguished from the background for each single time bin (i.e. TS $\le$ 25), TS value of the 10th time bin reaches to 21. Then 6-month and 3-month time bin light curves are further extracted. Intriguingly, in the time interval of MJD 58217 to MJD 58308 there is a $\gamma$-ray signal with ${\rm TS}\sim 30$. We note that rising of TS value of the target could be caused by flaring of the bright neighbors. Therefore, {\bf corresponding} light curves of two nearby strong background $\gamma$-ray sources, 4FGL~J1454.4+5124 and 4FGL~J1543.0+6130 (both 6$\degr$ away) whose TS values from the global fit are 8000 and 19000 respectively, are also extracted. Luckily, in this special epoch, no coincident $\gamma$-ray flares from the neighbors are found, also see Figure \ref{Fig.glc}. A residual TS map confirms the emergence of a valid $\gamma$-ray source towards GB 1508+5714 then, see Figure \ref{Fig.tsmap}. The optimum location of this source is constrained as R.A. 227.368$\degr$ and DEC. 57.111$\degr$ with a 95\% C. L. error radius of 0.3$\degr$ that embraces the radio position of GB 1508+5714. The source is spectrally soft, $\Gamma$ = 2.97 $\pm$ 0.25, and no significant spectral hardening is found compared with the global fit. But the flux then, $\rm (3.4 \pm 0.8)\times10^{-8}$ ph $\rm cm^{-2}$ $\rm s^{-1}$, is roughly four times of the 11.4-yr averaged value, $\rm (7.6 \pm 1.2)\times10^{-9}$ ph $\rm cm^{-2}$ $\rm s^{-1}$. Due to the limited statistics, it is impossible to obtain any variability information at timescale of a few days. Nevertheless, if the time range of the data narrows down to about 48 days (i.e. from MJD 58241 to MJD 58290), TS value of the source is still as large as 26. We also calculate the variability index \citep{2012ApJS..199...31N} value based on the 3-month time bin light curve here. The $\sigma_{var}$ increases from 3.0 \citep{2018ApJ...853..159L} to 4.3 after embracing the latest 2.4-yr $Fermi$-LAT data.

\subsection{iPTF and ZTF light curve data}
Here we adopt the light curve data archived by the IRSA\footnote{https://irsa.ipac.caltech.edu} from both iPTF and ZTF which use the 48 inch Schmidt telescope at the Palomar Observatory. The total field of view of iPTF is 7.26 $\rm deg^{2}$ and the depth of a single snapshot is R $\simeq$ 20.5 mag or $g$ $\simeq$ 21 mag when the exposure time is 60~sec under the median seeing (i.e. 2$\arcsec$) \citep{2016PASP..128k4502C}. In 2017, iPTF had transitioned to ZTF which has an enhanced 47 $\rm deg^{2}$ field and scans the northern sky at rates of $\sim$ 3760 $\rm deg^{2}$ per hour \citep{2019PASP..131a8002B,2019PASP..131a8003M,2019PASP..131g8001G}. With exposures in 30~sec, the median depths are $g \sim$ 20.8 mag and $r$ $\sim$ 20.6 mag (AB, 5$\sigma$). Additionally, $i$-band observations are also carried out in ZTF. 

\begin{figure}
\centering
\includegraphics[scale=0.65]{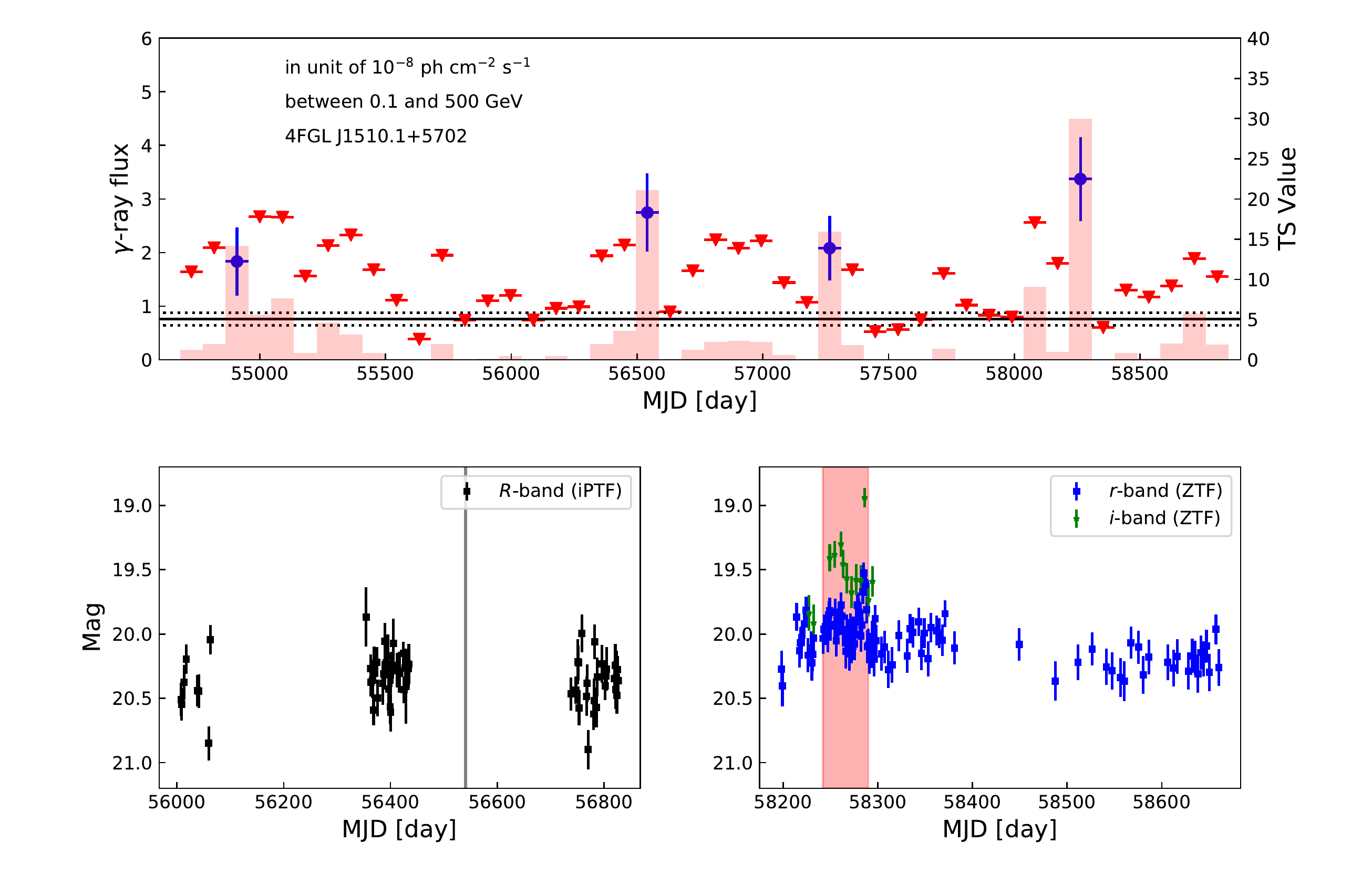}
\caption{$\gamma$-ray and optical long-term light curves of GB 1508+5714. {\it Upper} panel: the 3-month time bin $\gamma$-ray light curve, also shown in Figure \ref{Fig.glc}. The red shadow area in the bottom right panel corresponds to the 48-day time epoch when TS value of the target is about 26. The grey vertical line in the {\it Bottom Left} {\bf panel} marks the time when the TS value of one time bin in the $\gamma$-ray light curve reaches to $\sim$ 20.}
\label{Fig.mlc}
\end{figure}

\begin{figure}
\centering
\includegraphics[scale=0.8]{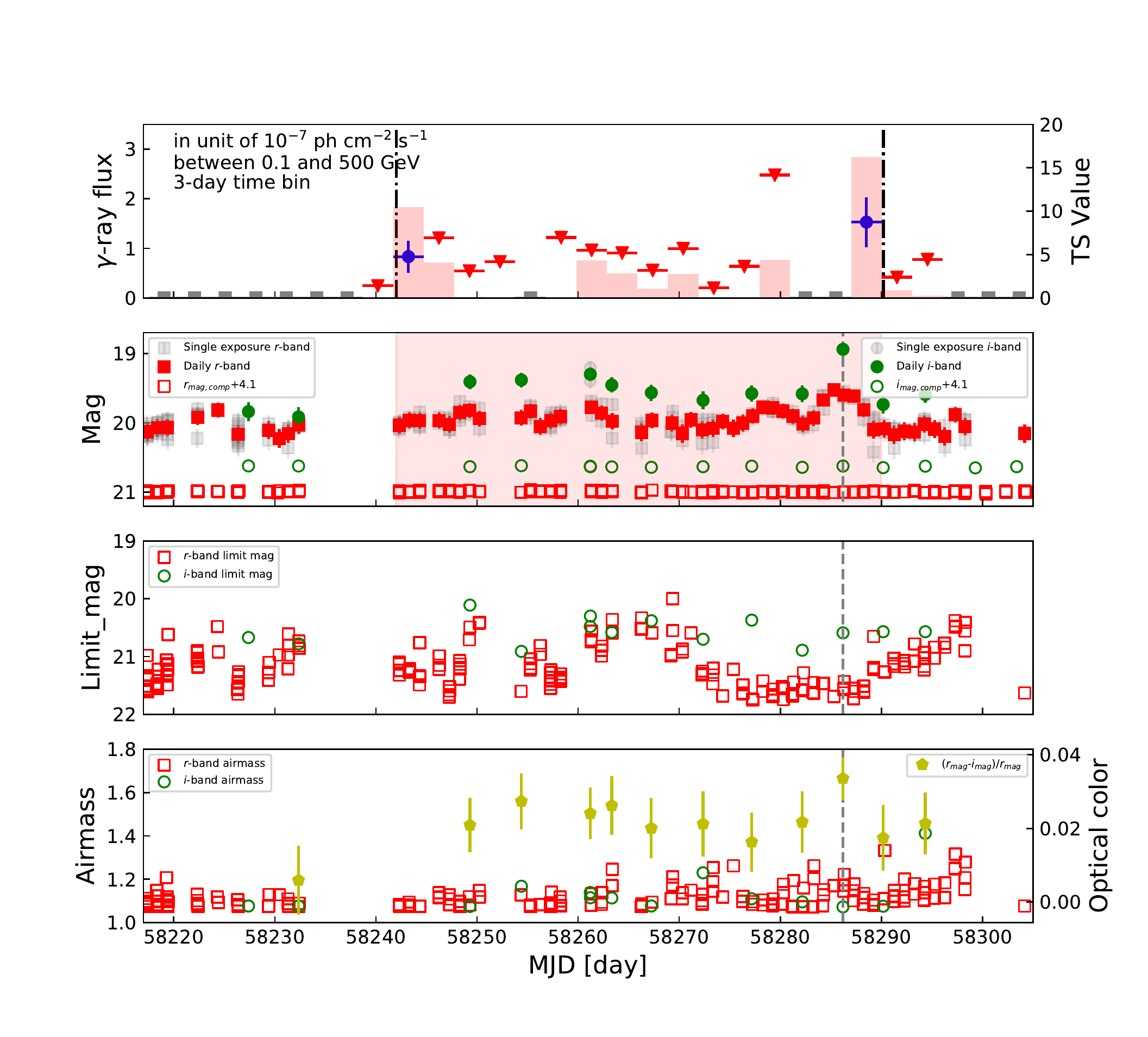}
\caption{3-day time bin $\gamma$-ray light curve and the zoomed-in ZTF light curve with each exposure as well as the corresponding limit mag and airmass. The grey squares in the {\it Upper} panel represent time epochs without valid {\it Fermi}-LAT exposure then. The red shadow area (also marked by the vertical dash-dotted lines in the {\it Upper} panel) represents the same 48-day time epoch that also shown in Figure \ref{Fig.mlc}. The averaged magnitudes of five comparison stars (i.e. the hollow points) are also plotted in the ZTF light curve panel. The vertical dashed line marks the time (i.e. MJD 58286.3) when the $i$-band flux density reaches to its peak value. The yellow pentagrams are optical spectral colors scaled by $r_{mag}$ corresponding to different epochs.}
\label{Fig.zlc}
\end{figure}

Recently, the second ZTF public data release (DR2) have been announced\footnote{https://www.ztf.caltech.edu/news/public-data-release-2}. Initially, we have looked up the co-added reference images. GB 1508+5714 is faint in the $g$-band image ($\simeq$ 22.1 mag, \citealt{2005AJ....130..367S}) but significantly detected in the frames of the rest two filters. Therefore, light curves in $r$-band and $i$-band for objects falling within a 5$\arcsec$ radius from position of the target are derived. There are in total 14 $i$-band exposures (in 13 days from MJD 58227 to MJD 58290) and 252 $r$-band exposures (in 104 days from MJD 58200 to MJD 58660). Only ZTF frames satisfied with {\tt catflags = 0} are selected. In addition, there are 160 $R$-band iPTF exposures (in 65 days from MJD 56008 to MJD 56825). The long-term daily averaged optical light curves are shown in Figure \ref{Fig.mlc}. Besides the daily optical light curves, a zoomed-in frame of the ZTF light curves in year 2018 with each exposure exhibited is also shown, together with the corresponding airmass and the limit mag, see Figure \ref{Fig.zlc}. Note that photometric zero-point corrections have been applied for the ZTF and iPTF light curves. Nevertheless, five comparison stars ($r_{mag} \sim$ 17) locating within 10$\arcmin$ from the position of GB 1508+5714 among the PTF Photometric Calibrator Catalog \citep{2012PASP..124..854O} are selected. The standard deviation of their ZTF light curves is small, less than 0.02 mag in the both $r$- and $i$-bands. The averaged magnitudes of the comparison stars are also plotted, together with the zoomed-in light curve of the target, see Figure \ref{Fig.zlc}. The typical mag of GB 1508+5714 for each single ZTF $r$-band exposure is $\sim$ 20 mag. But generally, the $r$-band flux in year 2018 ($r_{mean}$ $\simeq$ 20 mag) appears to be slightly brighter than that in year 2019 ($r_{mean}$ $\simeq$ 20.2 mag). One attractive feature in the $r$-band light curve is the existence of flares in June 2018. There is a 0.5 mag brightening from MJD 58282.2 ($r$ = 20.0 $\pm$ 0.12 mag) to MJD 58285.3 ($r$ = 19.5 $\pm$ 0.08 mag), and the source became fainter in 4 days ($r$ = 20.1 $\pm$ 0.13 mag at MJD 58289.2). The ascent phase $r$-band light curve gives a tight constraint on the doubling timescale at the source frame, $\tau_{doub, source} = \Delta t\times$ $\rm ln 2/ln(F_{1}/F_{2})/$(1+$z$) $\sim$ 0.9~day. Before this major flare, there is a minor flare peaking at MJD 58279.3 with brightening of about 0.3 mag, see Figure \ref{Fig.zlc}. On the other hand, though the $i$-band light curve is sparsely sampled, the $i$-band mag at MJD 58286.3 reached to 18.9 $\pm$ 0.08 mag, which is significantly brighter than that in MJD 58272.4 ($i$ = 19.7 $\pm$ 0.13 mag). A quantity defined as $\Delta mag/\sqrt{(magerr^{2}_{1}+magerr^{2}_{2})}$ is used to qualify the significance of variation between two photometric estimations. So the jump in $i$-band light curve suggest a significance level of $\sim 5\sigma$ of variability. Meanwhile, three $r$-band data points maintained at high flux state (i.e. $\sim 19.5$ mag) around MJD 58286, in contrast to the data at MJD 58282.2, which gives variability at the significance level of 5.5$\sigma$. A null hypothesis that all these flux variations are due to random fluctuation has been rejected with a significance level of 7.8$\sigma$. We would like to also remind that these observations were carried out under good conditions (i.e. $i_{\rm limit} \simeq$ 20.6 mag/$r_{\rm limit} \simeq$ 21.3 mag and airmass $<$ 1.3) and these optical emission variations can not be attributed to the bad weather or the poor air masses. We conclude that the optical variations in the ZTF light curves around MJD 58286 are robust, and GB 1508+571 was indeed undergoing an active phase then.  For the iPTF $R$-band light curve, as discussed in \cite{2018ApJ...853..159L}, no significant optical variability was found. 

\subsection{Implications of $\gamma$-ray and optical variability}
The 3-month $\gamma$-ray light curve of 4FGL J1510.1+5702 reveals that it is at high flux state in an epoch of several tens of days in year 2018, meanwhile, optical flux densities of GB 1508+5714 in two bands rise in the same epoch. To further investigate the relationship between these two domains of emissions, a 3-day time bin $\gamma$-ray light curve is presented, together with the zoomed-in ZTF light curves, see Figure \ref{Fig.zlc}. In spite of the limited statistics and no {\it Fermi}-LAT observation towards the target at the exact time of the optical flares, the time bin (i.e. centered at 58288.5) with the largest TS value in the 3-day $\gamma$-ray light curve is very close to the peaking time of the optical flares. Since optical variations are likely from the jet because of the large variability amplitude, based on the simultaneous $\gamma$-ray and optical brightening, we conclude that 4FGL J1510.1+5702 is the $\gamma$-ray counterpart of GB 1508+5714. There are three other time bins in the 3-month $\gamma$-ray light curve with the TS values $\geq$ 10, centered at MJD 54909, 57265 and 56540, respectively. The first two time intervals, they do not fall into the operation time range of iPTF/ZTF. Moreover, no iPTF/ZTF data of GB 1508+5714 around MJD 56540 are available, see Figure \ref{Fig.mlc}. Theoretically, in the leptonic radiation scenario, the optical and GeV $\gamma$-ray emissions of low synchrotron peaked blazars, including FSRQs, are proposed to be from the same population of emitting electrons. It is supported by correlated optical/$\gamma$-ray flares in FSRQs \citep[e.g.,][]{2010Natur.463..919A,2012ApJ...756...13B}. Meanwhile, the redder-when-brighter spectral variability behavior has been detected in the optical wavelengths of $\gamma$-ray FSRQs, which is explained by the influence of the blue and slowly varying accretion disk emission \citep[e.g.,][]{2012ApJ...756...13B,2018ApJ...856...80F}. For GB 1508+5714, a similar trend is shown. The optical spectral color, $r_{mag}$-$i_{mag}$, is 0.66 mag in MJD 58286, while it is 0.43 mag in MJD 58272, despite the relatively large photometric uncertainties. The optical spectral color scaled by $r_{mag}$ corresponding to observations in different epochs are plotted, shown in Figure \ref{Fig.zlc}. And its $i$-band variability amplitude is larger than that in $r$-band. All these facts suggests that the contribution of the jet emission becomes significant at the optical wavelengths of GB 1508+5714 when the jet activity is intense with rising $\gamma$-ray emission. Moreover, the significant $\gamma$-ray emission together with the rapid optical variation in $r$-band provide information of the emitting jet blob. The radius of emitting blob is constrained by the variability timescale, $R_{j}^{\prime}\leq$ c$\tau_{doub, source} \delta$, where $\tau_{doub, source} \sim$ 0.9 day for the current event. Meanwhile, assuming that the optical and $\gamma$-ray photons of GB 1508+5714 are from the same region, to avoid serious absorption on $\gamma$ rays from soft photons via $\gamma$$\gamma$ process, the corresponding optical depth should not be high,
\begin{equation}
\tau_{\gamma\gamma}(x^{\prime})=\frac{\sigma_{\rm T}}{5}n^{\prime}(x^{\prime}_{\rm t})x^{\prime}_{\rm t}R^{\prime}\lesssim 1,
\end{equation}
where $\sigma_{\rm T}$ is the scattering Thomson cross section, $n^{\prime}(x^{\prime})$ is the differential comoving number density of the target photon per energy, $x^{\prime}_{\rm t}$ is the energy of the target photon in dimensionless units, and $R^{\prime}$ is the absorption length \citep{1995MNRAS.273..583D,2008MNRAS.384L..19B}. The soft photons from jet itself could be responsible for the absorption. Since the highest energy of the detected $\gamma$-ray photons of GB 1508+5714 is $\sim$ 8~GeV, the corresponding soft photons are those detected at a few keV, (3$\times 10^{46}$ erg $\rm s^{-1}$, \citealt{2020ApJ...889..164M}), and the absorption length can be set as same as the radius of the emitting blob. A constraint of the Doppler factor of the jet blob, $\delta \gtrsim$ 7, is given.

\section{Discussions and Summary} \label{sec:discu}
Activity of $\gamma$-ray emissions of high-redshift ($z \geq$ 2) blazars is intense. Their peaking $\gamma$-ray luminosity is capable to reach to $>$ $\rm 10^{50}$ erg $\rm s^{-1}$ \citep{2015ApJ...799..143A,2016MNRAS.455.1881D}. Meanwhile, the variability amplitude can be as high as over one order of magnitude and the timescale of fast variations down to a few hours in the source frame have been often detected \citep[e.g.,][]{2013A&A...556A..71A,2015ApJ...799..143A,2018ApJ...853..159L}. By comparison, the $\gamma$-ray variation of GB 1508+5714 appears to be mild and the peaking luminosity is $\rm 2\times10^{48}$ erg $\rm s^{-1}$. Nevertheless, intraday optical variation of GB 1508+5714 is detected during the $\gamma$-ray brightening epoch. Considering intraday variability at optical/IR wavelengths detected in other high-redshift $\gamma$-ray blazars \citep[e.g.,][]{2018ApJ...853..159L,2019ApJ...879L...9L}, it is reasonable that highly beamed sources are more likely to be seen there due to the Malmquist bias. More importantly, the simultaneous brightening in optical (both in $r$- and $i$-bands) as well as $\gamma$ rays, provides the crucial evidence for GB 1508+5714 being a $\gamma$-ray emitter. GB 1508+5714 is a very important target for future multiwavelength campaign to probe the jet properties at high redshifts. It is worthwhile to note that there is another interesting source, NVSS J163547+362930 ($z$ = 3.6, \citealt{2014A&A...563A..54P}), from which detections of significant optical and $\gamma$-ray flares, though not simultaneously, have been reported \citep{2018ApJ...853..159L}. 

Besides blazars, $\gamma$-ray bursts (GRBs) are also strong extragalactic $\gamma$-ray emitters. The most distant GRB with GeV $\gamma$-ray detection by {\it Fermi}-LAT so far is GRB 080916C ($z_{\rm ph}$ = 4.35, \citealt{2009A&A...498...89G,2009Sci...323.1688A}). For blazars, the most distant source in hard X rays among the 105 month Swift-BAT all sky survey \citep{2018ApJS..235....4O} is B3 1428+422 ($z$ = 4.7, \citealt{1998MNRAS.294L...7H}).  Meanwhile, individual studies reveal several blazars (candidates) beyond redshift 5, including Q0906+6930 ($z$ = 5.48, \citealt{2004ApJ...610L...9R}), SDSS J102623.61+254259.5 ($z$ = 5.2, \citealt{2012MNRAS.426L..91S}), SDSS J013127.34$-$032100.1 ($z$ = 5.18, \citealt{2014ApJ...795L..29Y}, and SDSS J114657.79+403708.6 ($z$ = 5.0; \citealt{2014MNRAS.440L.111G}). Attempts that aim to break the redshift record by GRB 080916C have been made. A new spectrally soft transient (in an epoch of 10 months) $\gamma$-ray source (global significance of 4.1$\sigma$) towards B3 1428+422 is detected by {\it Fermi}-LAT, despite there is no significant excess there from a entire 9-year data averaged perspective \citep{2018ApJ...865L..17L}. Unfortunately, there are no available simultaneous observations that can be used to pin down the relationship between the transient $\gamma$-ray source and B3 1428+422. Meanwhile, TS value of one time bin of the $\gamma$-ray light curve of Q0906+6930 is about 12, locally $\sim 2\sigma$ \citep{2018ApJ...856..105A}. If blazars of $z \gtrsim$ 5 are as violently variable as ones of $z \simeq$ 3, their $\gamma$-ray emissions would be likely detected by {\it Fermi}-LAT during the flaring epochs. Moreover, the next generation MeV $\gamma$-ray all-sky surveyor, like the All sky Medium Energy Gamma-ray Observatory \citep{2020AAS...23537215M}, will play an important role of detecting high-redshift blazars. On the other hand, complementary time domain observations in other windows of the electromagnetic radiation are also crucial. Note that the optical emission of blazars at $z \gtrsim$ 5 is very faint, for example $R_{\rm Q0906+6930}$ = 21.7 mag \citep{2008ApJS..175...97H}. Upcoming 2-meter class wide-deep-fast optical sky surveyors in the northern hemisphere, like the Wide Field Survey Telescope \citep{2016SPIE10154E..2AL}, as well as the Large Synoptic Survey Telescope \citep{2019ApJ...873..111I}, will bring a bright future of catching activities from the jets in AGNs in the early cosmic time.

\begin{figure}
\centering
\includegraphics[scale=0.8]{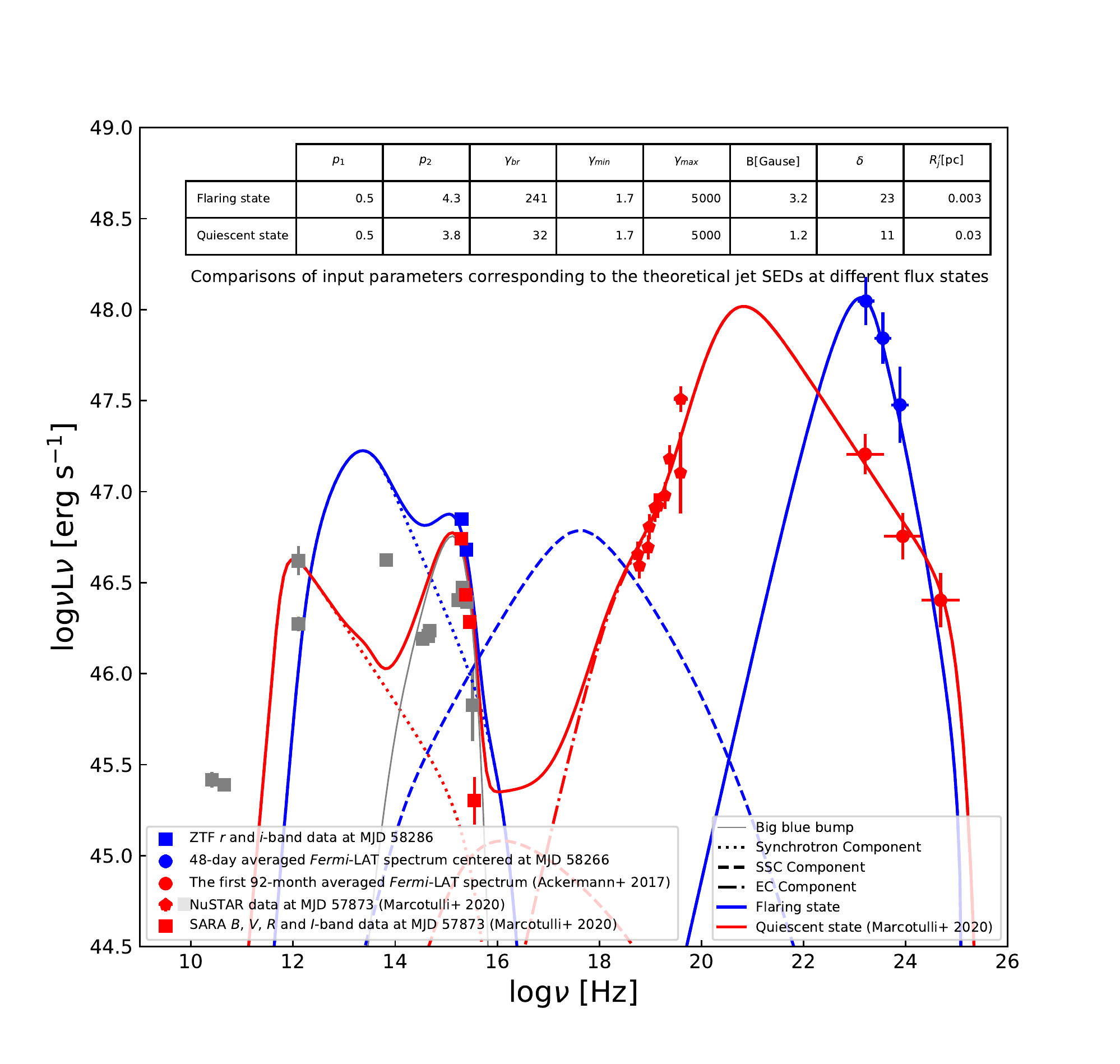}
\caption{SEDs in different flux states of GB 1508+5714 along with the theoretical descriptions. The color grey data points are un-simultaneous data and the grey line represents the accretion disk emission with $L_{d} = 7\times 10^{46}$ erg $\rm s^{-1}$ ($\sim 0.09 L_{edd}$) and $M_{BH} = 6.5\times 10^{9}M_{\odot}$ \citep{2020ApJ...889..164M}. The color blue points and lines correspond to the flaring state, while the red ones represent the quiescent state \citep{2020ApJ...889..164M}, respectively. The parameters corresponding to the Quiescent state SED are derived from \cite{2020ApJ...889..164M}. $\rm p_{1,2}$ are the indexes of the broken power-law radiative electron distribution; $\gamma_{br}$, $\gamma_{min}$ and $\gamma_{max}$ are the break, the minimum and maximum energies of the electron distribution, respectively; B is the magnetic field strength; $\delta$ is the Doppler boosting factor and $R_{j}^{\prime}$ is the radius of the emission blob. \tiny}
\label{Fig.sed}
\end{figure}

Finally, broadband SEDs of GB 1508+5714 corresponding to different flux states are presented, see Figure \ref{Fig.sed}. The data and the theoretical description of the quiescent SED are derived from \cite{2020ApJ...889..164M}, including data from simultaneous observations of Southeastern Association for Research in Astronomy's optical telescopes (SARA) and Nuclear Spectroscopic Telescope Array ({\it NuSTAR}) at MJD 57873, as well as the first 92-month averaged $\gamma$-ray spectrum detected by {\it Fermi}-LAT \citep{2017ApJ...837L...5A}. On the other hand, the high flux state SED consists of the ZTF $r$ and $i$-band fluxes at MJD 58286 together with a 48-day averaged $\gamma$-ray spectrum centered at MJD 58266 from {\it Fermi}-LAT. The classic single-zone homogeneous leptonic model embracing the synchrotron and IC processes (both SSC and EC) is used to describe the high flux state SED, in which the synchrotron self-absorption process and the Klein$-$Nishina effect in the IC scattering are considered. The jet emission is from a relativistic compact blob with a radius of $R_{j}^{\prime}$ embedded in the magnetic field and the external photon field. The transformations of frequency and luminosity between the jet frame and the observational frame are $\nu = \delta\nu^{\prime}/(1+z)$ and $\nu L_{\nu} = \delta^{4}\nu^{\prime}L^{\prime}_{\nu^{\prime}}$. We assume that the emitting electrons follow a broken power-law distribution,
\begin{equation}
N(\gamma ) \propto \left\{ \begin{array}{ll}
                    \gamma ^{-p_1}  &  \mbox{ $\gamma_{\rm min}\leq \gamma \leq \gamma_{br}$} \\
            \gamma _{\rm br}^{p_2-p_1} \gamma ^{-p_2}  &  \mbox{ $\gamma _{\rm br}<\gamma\leq\gamma_{\rm max}$, }
           \end{array}
       \right.
\label{Ngamma}
\end{equation}
where $\rm \gamma_{br}$, $\rm \gamma _{min}$ and $\rm \gamma _{max}$ are the break, the minimum and maximum energies of the electrons, and $p_{1,2}$ are indices of the broken power-law particle distribution. In the EC process, Ly$\alpha$ line emission is adopted as the origin of the external photon field. We extract the monochrome UV flux at 1350~\AA~and the luminosity of the Ly$\alpha$ line by analyzing the archival SDSS optical spectrum of GB 1508+5714. Therefore, the scale of the broad line region (BLR) can be inferred as $\rm r_{BLR} \sim$ 0.29~pc \citep{2015ApJ...801....8K} and hence the energy density of the external soft photons can be estimated as $\rm U_{ext} = L_{Ly\alpha}/4\pi r_{BLR}^{2}c \simeq 8\times10^{-3}$ erg $\rm cm^{-3}$. Since there is no simultaneous X-ray observations for the high flux state SED, we set the value of $p_{1}$ as 0.5, consistent with the value in \cite{2020ApJ...889..164M}, which is used to explain the extraordinary hard {\it NuSTAR} spectrum. As shown in Figure \ref{Fig.sed}, the leptonic scenario well reproduces high flux state SED and the corresponding input parameters are summarized. One of the major differences of the input parameter between SEDs at different flux states is the radius of the emitting blob. The rapid optical variation indicates a compact emitting region for the flaring epoch. If a conical jet geometry is assumed, the distance between the location of jet energy dissipation and the central SMBH then can be estimated as $\sim$ 0.1~pc which is smaller than it for the quiescent SED ($\sim$ 0.3~pc, \citealt{2020ApJ...889..164M}), and hence it is acceptable that the intensity of the magnetic field is higher for the former case. Meanwhile, significant enhances of the Doppler factor together with the $\rm \gamma_{br}$ that might be caused by ejecta of a new jet blob provide a natural explanation of the brightening of optical and $\gamma$-ray emissions of GB 1508+5714. In fact, the moving of SED bump peaks of FSRQs to shorter wavelengths during their flaring epoch is frequently detected \citep[e.g.,][]{2013MNRAS.432L..66G,2013ApJ...767....8Z,2015ApJ...807...79H}. For the high-redshift blazars, evolution of broadband SEDs of CGRaBS J0733+0456 has been also exhibited and theoretically described \citep{2019ApJ...879L...9L}, from which the input parameters agree with ones presented here.

In summary, we perform an investigation of the $\gamma$-ray and optical variability properties of GB 1508+5714. TS value from the analysis of the entire 11.4-yr {\it Fermi}-LAT data is doubled compared with the value listed in 4FGL. The $\gamma$-ray source is indeed at a high flux state in an epoch of several tens of days in year 2018. The flux then is about four times of the flux from the global fit, and the corresponding TS value reaches to 30. Meanwhile, at the same time, significant rise of the optical fluxes, both in $i$- and $r$-bands, are found through the ZTF light curves. The sign of fast variation in $r$-band and the redder-when-brighter optical spectral variability are also detected then. In consideration of the $\gamma$-ray and optical brightening, GB 1508+5714 is strongly suggested to be the {\it first} identified $\gamma$-ray blazar beyond reshift 4. A broadband SED in the high flux state is theoretically interpreted and compared with one in the quiescent flux state \citep{2020ApJ...889..164M}. Future multiwavelength campaigns are urged to further investigate its jet properties.

\acknowledgments
We appreciate the instructive suggestions from the anonymous referee. Lea Marcotulli is appreciated for sharing the observational and theoretical SED data of GB 1508+5714. This research has made use of data obtained from the High Energy Astrophysics Science Archive Research Center (HEASARC), provided by NASA's Goddard Space Flight Center. This research has made use of the NASA/IPAC Infrared Science Archive, which is funded by the NASA and operated by the California Institute of Technology. This study use data based on observations obtained with the Samuel Oschin 48-inch Telescope at the Palomar Observatory as part of the iPTF and ZTF projects. ZTF is supported by the National Science Foundation under Grant No. AST-1440341 and a collaboration including Caltech, IPAC, the Weizmann Institute for Science, the Oskar Klein Center at Stockholm University, the University of Maryland, the University of Washington, Deutsches Elektronen-Synchrotron and Humboldt University, Los Alamos National Laboratories, the TANGO Consortium of Taiwan, the University of Wisconsin at Milwaukee, and Lawrence Berkeley National Laboratories. Operations are conducted by COO, IPAC, and UW.

This work was supported in part by NSFC under grants 11525313 (i.e., Funds for Distinguished Young Scholars) and 11703093.

\bibliographystyle{aasjournal}
\bibliography{refs}

\begin{thebibliography}{}
\expandafter\ifx\csname natexlab\endcsname\relax\def\natexlab#1{#1}\fi
\providecommand{\url}[1]{\href{#1}{#1}}
\providecommand{\dodoi}[1]{doi:~\href{http://doi.org/#1}{\nolinkurl{#1}}}
\providecommand{\doeprint}[1]{\href{http://ascl.net/#1}{\nolinkurl{http://ascl.net/#1}}}
\providecommand{\doarXiv}[1]{\href{https://arxiv.org/abs/#1}{\nolinkurl{https://arxiv.org/abs/#1}}}

\bibitem[{{Abdo} {et~al.}(2009){Abdo}, {Ackermann}, {Arimoto}, {Asano},
  {Atwood}, {Axelsson}, {Baldini}, {Ballet}, {Band}, {Barbiellini}, {Baring},
  {Bastieri}, {Battelino}, {Baughman}, {Bechtol}, {Bellardi}, {Bellazzini},
  {Berenji}, {Bhat}, {Bissaldi}, {Bland ford}, {Bloom}, {Bogaert}, {Bogart},
  {Bonamente}, {Bonnell}, {Borgland}, {Bouvier}, {Bregeon}, {Brez}, {Briggs},
  {Brigida}, {Bruel}, {Burnett}, {Burrows}, {Busetto}, {Caliandro}, {Cameron},
  {Caraveo}, {Casandjian}, {Ceccanti}, {Cecchi}, {Celotti}, {Charles},
  {Chekhtman}, {Cheung}, {Chiang}, {Ciprini}, {Claus}, {Cohen-Tanugi},
  {Cominsky}, {Connaughton}, {Conrad}, {Costamante}, {Cutini}, {DeKlotz},
  {Dermer}, {de Angelis}, {de Palma}, {Digel}, {Dingus}, {do Couto e Silva},
  {Drell}, {Dubois}, {Dumora}, {Edmonds}, {Evans}, {Fabiani}, {Farnier},
  {Favuzzi}, {Finke}, {Fishman}, {Focke}, {Frailis}, {Fukazawa}, {Funk},
  {Fusco}, {Gargano}, {Gasparrini}, {Gehrels}, {Germani}, {Giebels},
  {Giglietto}, {Giommi}, {Giordano}, {Glanzman}, {Godfrey}, {Goldstein},
  {Granot}, {Greiner}, {Grenier}, {Grondin}, {Grove}, {Guillemot}, {Guiriec},
  {Haller}, {Hanabata}, {Harding}, {Hayashida}, {Hays}, {Morata}, {Hoover},
  {Hughes}, {J{\'o}hannesson}, {Johnson}, {Johnson}, {Johnson}, {Johnson},
  {Kamae}, {Katagiri}, {Kataoka}, {Kavelaars}, {Kawai}, {Kelly}, {Kennea},
  {Kerr}, {Kippen}, {Kn{\"o}dlseder}, {Kocevski}, {Kocian}, {Komin},
  {Kouveliotou}, {Kuehn}, {Kuss}, {Land e}, {Landriu}, {Larsson}, {Latronico},
  {Lavalley}, {Lee}, {Lee}, {Lemoine-Goumard}, {Lichti}, {Longo}, {Loparco},
  {Lott}, {Lovellette}, {Lubrano}, {Madejski}, {Makeev}, {Marangelli},
  {Mazziotta}, {McBreen}, {McEnery}, {McGlynn}, {Meegan}, {M{\'e}sz{\'a}ros},
  {Meurer}, {Michelson}, {Minuti}, {Mirizzi}, {Mitthumsiri}, {Mizuno},
  {Moiseev}, {Monte}, {Monzani}, {Moretti}, {Morselli}, {Moskalenko}, {Murgia},
  {Nakamori}, {Nelson}, {Nolan}, {Norris}, {Nuss}, {Ohno}, {Ohsugi}, {Okumura},
  {Omodei}, {Orland o}, {Ormes}, {Ozaki}, {Paciesas}, {Paneque}, {Panetta},
  {Parent}, {Pelassa}, {Pepe}, {Perri}, {Pesce-Rollins}, {Petrosian},
  {Pinchera}, {Piron}, {Porter}, {Preece}, {Rain{\`o}}, {Ramirez-Ruiz},
  {Rando}, {Rapposelli}, {Razzano}, {Razzaque}, {Rea}, {Reimer}, {Reimer},
  {Reposeur}, {Reyes}, {Ritz}, {Rochester}, {Rodriguez}, {Roth}, {Ryde},
  {Sadrozinski}, {Sanchez}, {Sander}, {Parkinson}, {Scargle}, {Schalk},
  {Segal}, {Sgr{\`o}}, {Shimokawabe}, {Siskind}, {Smith}, {Smith}, {Spandre},
  {Spinelli}, {Stamatikos}, {Starck}, {Stecker}, {Steinle}, {Stephens},
  {Strickman}, {Suson}, {Tagliaferri}, {Tajima}, {Takahashi}, {Takahashi},
  {Tanaka}, {Tenze}, {Thayer}, {Thayer}, {Thompson}, {Tibaldo}, {Torres},
  {Tosti}, {Tramacere}, {Turri}, {Tuvi}, {Usher}, {van der Horst}, {Vigiani},
  {Vilchez}, {Vitale}, {von Kienlin}, {Waite}, {Williams}, {Wilson-Hodge},
  {Winer}, {Wood}, {Wu}, {Yamazaki}, {Ylinen}, {Ziegler}, {Fermi LAT
  Collaboration}, \& {Fermi GBM Collaboration}}]{2009Sci...323.1688A}
{Abdo}, A.~A., {Ackermann}, M., {Arimoto}, M., {et~al.} 2009, Science, 323,
  1688, \dodoi{10.1126/science.1169101}

\bibitem[{Abdo {et~al.}(2010)}]{2010Natur.463..919A}
Abdo, A.~A., {et~al.} 2010, Nature, 463, 919, \dodoi{10.1038/nature08841}

\bibitem[{{Abdo} {et~al.}(2015){Abdo}, {Ackermann}, {Ajello}, {Allafort},
  {Amin}, {Baldini}, {Barbiellini}, {Bastieri}, {Bechtol}, {Bellazzini}, {Bland
  ford}, {Bonamente}, {Borgland}, {Bregeon}, {Brigida}, {Buehler}, {Bulmash},
  {Buson}, {Caliandro}, {Cameron}, {Caraveo}, {Cavazzuti}, {Cecchi}, {Charles},
  {Cheung}, {Chiang}, {Chiaro}, {Ciprini}, {Claus}, {Cohen-Tanugi}, {Conrad},
  {Corbet}, {Cutini}, {D'Ammando}, {de Angelis}, {de Palma}, {Dermer}, {Drell},
  {Drlica-Wagner}, {Favuzzi}, {Finke}, {Focke}, {Fukazawa}, {Fusco}, {Gargano},
  {Gasparrini}, {Gehrels}, {Giglietto}, {Giordano}, {Giroletti}, {Glanzman},
  {Grenier}, {Grove}, {Guiriec}, {Hadasch}, {Hayashida}, {Hays}, {Hughes},
  {Inoue}, {Jackson}, {Jogler}, {J{\'o}hannesson}, {Johnson}, {Kamae},
  {Kn{\"o}dlseder}, {Kuss}, {Lande}, {Larsson}, {Latronico}, {Longo},
  {Loparco}, {Lott}, {Lovellette}, {Lubrano}, {Madejski}, {Mazziotta},
  {Mehault}, {Michelson}, {Mizuno}, {Monzani}, {Morselli}, {Moskalenko},
  {Murgia}, {Nemmen}, {Nuss}, {Ohno}, {Ohsugi}, {Paneque}, {Perkins},
  {Pesce-Rollins}, {Piron}, {Pivato}, {Porter}, {Rain{\`o}}, {Rando},
  {Razzano}, {Reimer}, {Reimer}, {Reyes}, {Ritz}, {Romoli}, {Roth}, {Saz
  Parkinson}, {Sgr{\`o}}, {Siskind}, {Spandre}, {Spinelli}, {Takahashi},
  {Takeuchi}, {Tanaka}, {Thayer}, {Thayer}, {Thompson}, {Tibaldo}, {Tinivella},
  {Torres}, {Tosti}, {Troja}, {Tronconi}, {Usher}, {Vand enbroucke},
  {Vasileiou}, {Vianello}, {Vitale}, {Waite}, {Werner}, {Winer}, \&
  {Wood}}]{2015ApJ...799..143A}
{Abdo}, A.~A., {Ackermann}, M., {Ajello}, M., {et~al.} 2015, \apj, 799, 143,
  \dodoi{10.1088/0004-637X/799/2/143}

\bibitem[{{Abdollahi} {et~al.}(2020){Abdollahi}, {Acero}, {Ackermann},
  {Ajello}, {Atwood}, {Axelsson}, {Baldini}, {Ballet}, {Barbiellini},
  {Bastieri}, {Becerra Gonzalez}, {Bellazzini}, {Berretta}, {Bissaldi}, {Bland
  ford}, {Bloom}, {Bonino}, {Bottacini}, {Brandt}, {Bregeon}, {Bruel},
  {Buehler}, {Burnett}, {Buson}, {Cameron}, {Caputo}, {Caraveo}, {Casandjian},
  {Castro}, {Cavazzuti}, {Charles}, {Chaty}, {Chen}, {Cheung}, {Chiaro},
  {Ciprini}, {Cohen-Tanugi}, {Cominsky}, {Coronado-Bl{\'a}zquez}, {Costantin},
  {Cuoco}, {Cutini}, {D'Ammando}, {DeKlotz}, {Torre Luque}, {de Palma},
  {Desai}, {Digel}, {Lalla}, {Mauro}, {Venere}, {Dom{\'\i}nguez}, {Dumora},
  {Dirirsa}, {Fegan}, {Ferrara}, {Franckowiak}, {Fukazawa}, {Funk}, {Fusco},
  {Gargano}, {Gasparrini}, {Giglietto}, {Giommi}, {Giordano}, {Giroletti},
  {Glanzman}, {Green}, {Grenier}, {Griffin}, {Grondin}, {Grove}, {Guiriec},
  {Harding}, {Hayashi}, {Hays}, {Hewitt}, {Horan}, {J{\'o}hannesson},
  {Johnson}, {Kamae}, {Kerr}, {Kocevski}, {Kovac'evic'}, {Kuss}, {Landriu},
  {Larsson}, {Latronico}, {Lemoine-Goumard}, {Li}, {Liodakis}, {Longo},
  {Loparco}, {Lott}, {Lovellette}, {Lubrano}, {Madejski}, {Maldera},
  {Malyshev}, {Manfreda}, {Marchesini}, {Marcotulli}, {Mart{\'\i}-Devesa},
  {Martin}, {Massaro}, {Mazziotta}, {McEnery}, {Mereu}, {Meyer}, {Michelson},
  {Mirabal}, {Mizuno}, {Monzani}, {Morselli}, {Moskalenko}, {Negro}, {Nuss},
  {Ojha}, {Omodei}, {Orienti}, {Orlando}, {Ormes}, {Palatiello}, {Paliya},
  {Paneque}, {Pei}, {Pe{\~n}a-Herazo}, {Perkins}, {Persic}, {Pesce-Rollins},
  {Petrosian}, {Petrov}, {Piron}, {Poon}, {Porter}, {Principe}, {Rain{\`o}},
  {Rando}, {Razzano}, {Razzaque}, {Reimer}, {Reimer}, {Remy}, {Reposeur},
  {Romani}, {Parkinson}, {Schinzel}, {Serini}, {Sgr{\`o}}, {Siskind}, {Smith},
  {Spandre}, {Spinelli}, {Strong}, {Suson}, {Tajima}, {Takahashi}, {Tak},
  {Thayer}, {Thompson}, {Tibaldo}, {Torres}, {Torresi}, {Valverde}, {Klaveren},
  {Zyl}, {Wood}, {Yassine}, \& {Zaharijas}}]{2020ApJS..247...33A}
{Abdollahi}, S., {Acero}, F., {Ackermann}, M., {et~al.} 2020, \apjs, 247, 33,
  \dodoi{10.3847/1538-4365/ab6bcb}

\bibitem[{{Ackermann} {et~al.}(2017){Ackermann}, {Ajello}, {Baldini}, {Ballet},
  {Barbiellini}, {Bastieri}, {Becerra Gonzalez}, {Bellazzini}, {Bissaldi},
  {Blandford}, {Bloom}, {Bonino}, {Bottacini}, {Bregeon}, {Bruel}, {Buehler},
  {Buson}, {Cameron}, {Caragiulo}, {Caraveo}, {Cavazzuti}, {Cecchi}, {Cheung},
  {Chiang}, {Chiaro}, {Ciprini}, {Conrad}, {Costantin}, {Costanza}, {Cutini},
  {D'Ammando}, {de Palma}, {Desiante}, {Digel}, {Di Lalla}, {Di Mauro}, {Di
  Venere}, {Dom{\'\i}nguez}, {Drell}, {Favuzzi}, {Fegan}, {Ferrara}, {Finke},
  {Focke}, {Fukazawa}, {Funk}, {Fusco}, {Gargano}, {Gasparrini}, {Giglietto},
  {Giordano}, {Giroletti}, {Green}, {Grenier}, {Guillemot}, {Guiriec},
  {Hartmann}, {Hays}, {Horan}, {Jogler}, {J{\'o}hannesson}, {Johnson}, {Kuss},
  {La Mura}, {Larsson}, {Latronico}, {Li}, {Longo}, {Loparco}, {Lovellette},
  {Lubrano}, {Magill}, {Maldera}, {Manfreda}, {Marcotulli}, {Mazziotta},
  {Michelson}, {Mirabal}, {Mitthumsiri}, {Mizuno}, {Monzani}, {Morselli},
  {Moskalenko}, {Negro}, {Nuss}, {Ohsugi}, {Ojha}, {Omodei}, {Orienti},
  {Orlando}, {Ormes}, {Paliya}, {Paneque}, {Perkins}, {Persic},
  {Pesce-Rollins}, {Piron}, {Porter}, {Principe}, {Rain{\`o}}, {Rando}, {Rani},
  {Razzano}, {Razzaque}, {Reimer}, {Reimer}, {Romani}, {Sgr{\`o}}, {Simone},
  {Siskind}, {Spada}, {Spandre}, {Spinelli}, {Stalin}, {Stawarz}, {Suson},
  {Takahashi}, {Tanaka}, {Thayer}, {Thompson}, {Torres}, {Torresi}, {Tosti},
  {Troja}, {Vianello}, \& {Wood}}]{2017ApJ...837L...5A}
{Ackermann}, M., {Ajello}, M., {Baldini}, L., {et~al.} 2017, \apjl, 837, L5,
  \dodoi{10.3847/2041-8213/aa5fff}

\bibitem[{Akyuz {et~al.}(2013)Akyuz, Thompson, Donato, Perkins, Fuhrmann,
  Angelakis, Zensus, Larsson, Sokolovsky, \& Kurtanidze}]{2013A&A...556A..71A}
Akyuz, A., Thompson, D.~J., Donato, D., {et~al.} 2013, Astron. Astrophys., 556,
  A71, \dodoi{10.1051/0004-6361/201321721}

\bibitem[{{An} \& {Romani}(2018)}]{2018ApJ...856..105A}
{An}, H., \& {Romani}, R.~W. 2018, \apj, 856, 105,
  \dodoi{10.3847/1538-4357/aab435}

\bibitem[{Atwood {et~al.}(2009)}]{2009ApJ...697.1071A}
Atwood, W.~B., {et~al.} 2009, Astrophys. J., 697, 1071,
  \dodoi{10.1088/0004-637X/697/2/1071}

\bibitem[{{Begelman} {et~al.}(2008){Begelman}, {Fabian}, \&
  {Rees}}]{2008MNRAS.384L..19B}
{Begelman}, M.~C., {Fabian}, A.~C., \& {Rees}, M.~J. 2008, \mnras, 384, L19,
  \dodoi{10.1111/j.1745-3933.2007.00413.x}

\bibitem[{{Bellm} {et~al.}(2019){Bellm}, {Kulkarni}, {Graham}, {Dekany},
  {Smith}, {Riddle}, {Masci}, {Helou}, {Prince}, {Adams}, {Barbarino},
  {Barlow}, {Bauer}, {Beck}, {Belicki}, {Biswas}, {Blagorodnova}, {Bodewits},
  {Bolin}, {Brinnel}, {Brooke}, {Bue}, {Bulla}, {Burruss}, {Cenko}, {Chang},
  {Connolly}, {Coughlin}, {Cromer}, {Cunningham}, {De}, {Delacroix}, {Desai},
  {Duev}, {Eadie}, {Farnham}, {Feeney}, {Feindt}, {Flynn}, {Franckowiak},
  {Frederick}, {Fremling}, {Gal-Yam}, {Gezari}, {Giomi}, {Goldstein},
  {Golkhou}, {Goobar}, {Groom}, {Hacopians}, {Hale}, {Henning}, {Ho}, {Hover},
  {Howell}, {Hung}, {Huppenkothen}, {Imel}, {Ip}, {Ivezi{\'c}}, {Jackson},
  {Jones}, {Juric}, {Kasliwal}, {Kaspi}, {Kaye}, {Kelley}, {Kowalski},
  {Kramer}, {Kupfer}, {Landry}, {Laher}, {Lee}, {Lin}, {Lin}, {Lunnan},
  {Giomi}, {Mahabal}, {Mao}, {Miller}, {Monkewitz}, {Murphy}, {Ngeow},
  {Nordin}, {Nugent}, {Ofek}, {Patterson}, {Penprase}, {Porter}, {Rauch},
  {Rebbapragada}, {Reiley}, {Rigault}, {Rodriguez}, {van Roestel}, {Rusholme},
  {van Santen}, {Schulze}, {Shupe}, {Singer}, {Soumagnac}, {Stein}, {Surace},
  {Sollerman}, {Szkody}, {Taddia}, {Terek}, {Van Sistine}, {van Velzen},
  {Vestrand}, {Walters}, {Ward}, {Ye}, {Yu}, {Yan}, \&
  {Zolkower}}]{2019PASP..131a8002B}
{Bellm}, E.~C., {Kulkarni}, S.~R., {Graham}, M.~J., {et~al.} 2019, \pasp, 131,
  018002, \dodoi{10.1088/1538-3873/aaecbe}

\bibitem[{{Blandford} {et~al.}(2019){Blandford}, {Meier}, \&
  {Readhead}}]{2019ARA&A..57..467B}
{Blandford}, R., {Meier}, D., \& {Readhead}, A. 2019, \araa, 57, 467,
  \dodoi{10.1146/annurev-astro-081817-051948}

\bibitem[{{Blandford} \& {Rees}(1978)}]{1978bllo.conf..328B}
{Blandford}, R.~D., \& {Rees}, M.~J. 1978, in BL Lac Objects, ed. A.~M.
  {Wolfe}, 328--341

\bibitem[{{B{\l}a{\.z}ejowski} {et~al.}(2000){B{\l}a{\.z}ejowski}, {Sikora},
  {Moderski}, \& {Madejski}}]{2000ApJ...545..107B}
{B{\l}a{\.z}ejowski}, M., {Sikora}, M., {Moderski}, R., \& {Madejski}, G.~M.
  2000, \apj, 545, 107, \dodoi{10.1086/317791}

\bibitem[{{Bonning} {et~al.}(2012){Bonning}, {Urry}, {Bailyn}, {Buxton},
  {Chatterjee}, {Coppi}, {Fossati}, {Isler}, \&
  {Maraschi}}]{2012ApJ...756...13B}
{Bonning}, E., {Urry}, C.~M., {Bailyn}, C., {et~al.} 2012, \apj, 756, 13,
  \dodoi{10.1088/0004-637X/756/1/13}

\bibitem[{{Cao} {et~al.}(2016){Cao}, {Nugent}, \&
  {Kasliwal}}]{2016PASP..128k4502C}
{Cao}, Y., {Nugent}, P.~E., \& {Kasliwal}, M.~M. 2016, \pasp, 128, 114502,
  \dodoi{10.1088/1538-3873/128/969/114502}

\bibitem[{{D'Ammando} \& {Orienti}(2016)}]{2016MNRAS.455.1881D}
{D'Ammando}, F., \& {Orienti}, M. 2016, \mnras, 455, 1881,
  \dodoi{10.1093/mnras/stv2452}

\bibitem[{{Dermer} \& {Schlickeiser}(1993)}]{1993ApJ...416..458D}
{Dermer}, C.~D., \& {Schlickeiser}, R. 1993, \apj, 416, 458,
  \dodoi{10.1086/173251}

\bibitem[{{Dondi} \& {Ghisellini}(1995)}]{1995MNRAS.273..583D}
{Dondi}, L., \& {Ghisellini}, G. 1995, \mnras, 273, 583,
  \dodoi{10.1093/mnras/273.3.583}

\bibitem[{{Fan} {et~al.}(2018){Fan}, {Li}, {Liao}, {Chen}, {Liu}, {Lu}, {Yan},
  {Zhang}, {Guo}, {Wu}, \& {Bai}}]{2018ApJ...856...80F}
{Fan}, X.-L., {Li}, S.-K., {Liao}, N.-H., {et~al.} 2018, \apj, 856, 80,
  \dodoi{10.3847/1538-4357/aab09d}

\bibitem[{{$Fermi$-LAT Collaboration} {et~al.}(2018){$Fermi$-LAT
  Collaboration}, {Abdollahi}, {Ackermann}, {Ajello}, {Atwood}, {Baldini},
  {Ballet}, {Barbiellini}, {Bastieri}, {Becerra Gonzalez}, {Bellazzini},
  {Bissaldi}, {Blandford}, {Bloom}, {Bonino}, {Bottacini}, {Buson}, {Bregeon},
  {Bruel}, {Buehler}, {Cameron}, {Caputo}, {Caraveo}, {Cavazzuti}, {Charles},
  {Chen}, {Cheung}, {Chiaro}, {Ciprini}, {Cohen-Tanugi}, {Cominsky}, {Conrad},
  {Costantin}, {Cutini}, {D'Ammando}, {de Palma}, {Desai}, {Digel}, {Di Lalla},
  {Di Mauro}, {Di Venere}, {Dom{\'\i}nguez}, {Favuzzi}, {Fegan}, {Finke},
  {Franckowiak}, {Fukazawa}, {Funk}, {Fusco}, {Gallardo Romero}, {Gargano},
  {Gasparrini}, {Giglietto}, {Giordano}, {Giroletti}, {Green}, {Grenier},
  {Guillemot}, {Guiriec}, {Hartmann}, {Hays}, {Helgason}, {Horan},
  {J{\'o}hannesson}, {Kocevski}, {Kuss}, {Larsson}, {Latronico}, {Li}, {Longo},
  {Loparco}, {Lott}, {Lovellette}, {Lubrano}, {Madejski}, {Magill}, {Maldera},
  {Manfreda}, {Marcotulli}, {Mazziotta}, {McEnery}, {Meyer}, {Michelson},
  {Mizuno}, {Monzani}, {Morselli}, {Moskalenko}, {Negro}, {Nuss}, {Ojha},
  {Omodei}, {Orienti}, {Orlando}, {Ormes}, {Palatiello}, {Paliya}, {Paneque},
  {Perkins}, {Persic}, {Pesce-Rollins}, {Petrosian}, {Piron}, {Porter},
  {Primack}, {Principe}, {Rain{\`o}}, {Rand o}, {Razzano}, {Razzaque},
  {Reimer}, {Reimer}, {Saz Parkinson}, {Sgr{\`o}}, {Siskind}, {Spandre},
  {Spinelli}, {Suson}, {Tajima}, {Takahashi}, {Thayer}, {Tibaldo}, {Torres},
  {Torresi}, {Tosti}, {Tramacere}, {Troja}, {Valverde}, {Vianello}, {Vogel},
  {Wood}, \& {Zaharijas}}]{2018Sci...362.1031F}
{$Fermi$-LAT Collaboration}, {Abdollahi}, S., {Ackermann}, M., {et~al.} 2018,
  Science, 362, 1031, \dodoi{10.1126/science.aat8123}

\bibitem[{{Ghisellini} {et~al.}(2014{\natexlab{a}}){Ghisellini}, {Sbarrato},
  {Tagliaferri}, {Foschini}, {Tavecchio}, {Ghirlanda}, {Braito}, \&
  {Gehrels}}]{2014MNRAS.440L.111G}
{Ghisellini}, G., {Sbarrato}, T., {Tagliaferri}, G., {et~al.}
  2014{\natexlab{a}}, \mnras, 440, L111, \dodoi{10.1093/mnrasl/slu032}

\bibitem[{{Ghisellini} {et~al.}(2013){Ghisellini}, {Tavecchio}, {Foschini},
  {Bonnoli}, \& {Tagliaferri}}]{2013MNRAS.432L..66G}
{Ghisellini}, G., {Tavecchio}, F., {Foschini}, L., {Bonnoli}, G., \&
  {Tagliaferri}, G. 2013, \mnras, 432, L66, \dodoi{10.1093/mnrasl/slt041}

\bibitem[{{Ghisellini} {et~al.}(2014{\natexlab{b}}){Ghisellini}, {Tavecchio},
  {Maraschi}, {Celotti}, \& {Sbarrato}}]{2014Natur.515..376G}
{Ghisellini}, G., {Tavecchio}, F., {Maraschi}, L., {Celotti}, A., \&
  {Sbarrato}, T. 2014{\natexlab{b}}, \nat, 515, 376,
  \dodoi{10.1038/nature13856}

\bibitem[{{Ghisellini} {et~al.}(2010){Ghisellini}, {Della Ceca}, {Volonteri},
  {Ghirland a}, {Tavecchio}, {Foschini}, {Tagliaferri}, {Haardt}, {Pareschi},
  \& {Grindlay}}]{2010MNRAS.405..387G}
{Ghisellini}, G., {Della Ceca}, R., {Volonteri}, M., {et~al.} 2010, \mnras,
  405, 387, \dodoi{10.1111/j.1365-2966.2010.16449.x}

\bibitem[{{Graham} {et~al.}(2019){Graham}, {Kulkarni}, {Bellm}, {Adams},
  {Barbarino}, {Blagorodnova}, {Bodewits}, {Bolin}, {Brady}, {Cenko}, {Chang},
  {Coughlin}, {De}, {Eadie}, {Farnham}, {Feindt}, {Franckowiak}, {Fremling},
  {Gezari}, {Ghosh}, {Goldstein}, {Golkhou}, {Goobar}, {Ho}, {Huppenkothen},
  {Ivezi{\'c}}, {Jones}, {Juric}, {Kaplan}, {Kasliwal}, {Kelley}, {Kupfer},
  {Lee}, {Lin}, {Lunnan}, {Mahabal}, {Miller}, {Ngeow}, {Nugent}, {Ofek},
  {Prince}, {Rauch}, {van Roestel}, {Schulze}, {Singer}, {Sollerman}, {Taddia},
  {Yan}, {Ye}, {Yu}, {Barlow}, {Bauer}, {Beck}, {Belicki}, {Biswas}, {Brinnel},
  {Brooke}, {Bue}, {Bulla}, {Burruss}, {Connolly}, {Cromer}, {Cunningham},
  {Dekany}, {Delacroix}, {Desai}, {Duev}, {Feeney}, {Flynn}, {Frederick},
  {Gal-Yam}, {Giomi}, {Groom}, {Hacopians}, {Hale}, {Helou}, {Henning},
  {Hover}, {Hillenbrand}, {Howell}, {Hung}, {Imel}, {Ip}, {Jackson}, {Kaspi},
  {Kaye}, {Kowalski}, {Kramer}, {Kuhn}, {Landry}, {Laher}, {Mao}, {Masci},
  {Monkewitz}, {Murphy}, {Nordin}, {Patterson}, {Penprase}, {Porter},
  {Rebbapragada}, {Reiley}, {Riddle}, {Rigault}, {Rodriguez}, {Rusholme}, {van
  Santen}, {Shupe}, {Smith}, {Soumagnac}, {Stein}, {Surace}, {Szkody}, {Terek},
  {Van Sistine}, {van Velzen}, {Vestrand}, {Walters}, {Ward}, {Zhang}, \&
  {Zolkower}}]{2019PASP..131g8001G}
{Graham}, M.~J., {Kulkarni}, S.~R., {Bellm}, E.~C., {et~al.} 2019, \pasp, 131,
  078001, \dodoi{10.1088/1538-3873/ab006c}

\bibitem[{{Greiner} {et~al.}(2009){Greiner}, {Clemens}, {Kr{\"u}hler}, {von
  Kienlin}, {Rau}, {Sari}, {Fox}, {Kawai}, {Afonso}, {Ajello}, {Berger},
  {Cenko}, {Cucchiara}, {Filgas}, {Klose}, {K{\"u}pc{\"u} Yolda{\textcommabelow
  s}}, {Lichti}, {L{\"o}w}, {McBreen}, {Nagayama}, {Rossi}, {Sato}, {Szokoly},
  {Yolda{\textcommabelow s}}, \& {Zhang}}]{2009A&A...498...89G}
{Greiner}, J., {Clemens}, C., {Kr{\"u}hler}, T., {et~al.} 2009, \aap, 498, 89,
  \dodoi{10.1051/0004-6361/200811571}

\bibitem[{{Hayashida} {et~al.}(2015){Hayashida}, {Nalewajko}, {Madejski},
  {Sikora}, {Itoh}, {Ajello}, {Blandford}, {Buson}, {Chiang}, {Fukazawa},
  {Furniss}, {Urry}, {Hasan}, {Harrison}, {Alexand er}, {Balokovi{\'c}},
  {Barret}, {Boggs}, {Christensen}, {Craig}, {Forster}, {Giommi},
  {Grefenstette}, {Hailey}, {Hornstrup}, {Kitaguchi}, {Koglin}, {Madsen},
  {Mao}, {Miyasaka}, {Mori}, {Perri}, {Pivovaroff}, {Puccetti}, {Rana},
  {Stern}, {Tagliaferri}, {Westergaard}, {Zhang}, {Zoglauer}, {Gurwell},
  {Uemura}, {Akitaya}, {Kawabata}, {Kawaguchi}, {Kanda}, {Moritani}, {Takaki},
  {Ui}, {Yoshida}, {Agarwal}, \& {Gupta}}]{2015ApJ...807...79H}
{Hayashida}, M., {Nalewajko}, K., {Madejski}, G.~M., {et~al.} 2015, \apj, 807,
  79, \dodoi{10.1088/0004-637X/807/1/79}

\bibitem[{{Healey} {et~al.}(2008){Healey}, {Romani}, {Cotter}, {Michelson},
  {Schlafly}, {Readhead}, {Giommi}, {Chaty}, {Grenier}, \&
  {Weintraub}}]{2008ApJS..175...97H}
{Healey}, S.~E., {Romani}, R.~W., {Cotter}, G., {et~al.} 2008, \apjs, 175, 97,
  \dodoi{10.1086/523302}

\bibitem[{{Hook} \& {McMahon}(1998)}]{1998MNRAS.294L...7H}
{Hook}, I.~M., \& {McMahon}, R.~G. 1998, \mnras, 294, L7,
  \dodoi{10.1046/j.1365-8711.1998.01368.x}

\bibitem[{{Hook} {et~al.}(1995){Hook}, {McMahon}, {Patnaik}, {Browne},
  {Wilkinson}, {Irwin}, \& {Hazard}}]{1995MNRAS.273L..63H}
{Hook}, I.~M., {McMahon}, R.~G., {Patnaik}, A.~R., {et~al.} 1995, \mnras, 273,
  L63, \dodoi{10.1093/mnras/273.1.L63}

\bibitem[{{IceCube Collaboration} {et~al.}(2018){IceCube Collaboration},
  {Aartsen}, {Ackermann}, {Adams}, {Aguilar}, {Ahlers}, {Ahrens}, {Al Samarai},
  {Altmann}, {Andeen}, {Anderson}, {Ansseau}, {Anton}, {Arg{\"u}elles},
  {Auffenberg}, {Axani}, {Bagherpour}, {Bai}, {Barron}, {Barwick}, {Baum},
  {Bay}, {Beatty}, {Becker Tjus}, {Becker}, {BenZvi}, {Berley}, {Bernardini},
  {Besson}, {Binder}, {Bindig}, {Blaufuss}, {Blot}, {Bohm}, {B{\"o}rner},
  {Bos}, {B{\"o}ser}, {Botner}, {Bourbeau}, {Bourbeau}, {Bradascio}, {Braun},
  {Brenzke}, {Bretz}, {Bron}, {Brostean-Kaiser}, {Burgman}, {Busse}, {Carver},
  {Cheung}, {Chirkin}, {Christov}, {Clark}, {Classen}, {Coenders}, {Collin},
  {Conrad}, {Coppin}, {Correa}, {Cowen}, {Cross}, {Dave}, {Day}, {de
  Andr{\'e}}, {De Clercq}, {DeLaunay}, {Dembinski}, {De Ridder}, {Desiati}, {de
  Vries}, {de Wasseige}, {de With}, {DeYoung}, {D{\'\i}az-V{\'e}lez}, {di
  Lorenzo}, {Dujmovic}, {Dumm}, {Dunkman}, {Dvorak}, {Eberhardt}, {Ehrhardt},
  {Eichmann}, {Eller}, {Evenson}, {Fahey}, {Fazely}, {Felde}, {Filimonov},
  {Finley}, {Flis}, {Franckowiak}, {Friedman}, {Fritz}, {Gaisser}, {Gallagher},
  {Gerhardt}, {Ghorbani}, {Glauch}, {Gl{\"u}senkamp}, {Goldschmidt},
  {Gonzalez}, {Grant}, {Griffith}, {Haack}, {Hallgren}, {Halzen}, {Hanson},
  {Hebecker}, {Heereman}, {Helbing}, {Hellauer}, {Hickford}, {Hignight},
  {Hill}, {Hoffman}, {Hoffmann}, {Hoinka}, {Hokanson-Fasig}, {Hoshina},
  {Huang}, {Huber}, {Hultqvist}, {H{\"u}nnefeld}, {Hussain}, {In}, {Iovine},
  {Ishihara}, {Jacobi}, {Japaridze}, {Jeong}, {Jero}, {Jones}, {Kalaczynski},
  {Kang}, {Kappes}, {Kappesser}, {Karg}, {Karle}, {Katz}, {Kauer}, {Keivani},
  {Kelley}, {Kheirandish}, {Kim}, {Kim}, {Kintscher}, {Kiryluk}, {Kittler},
  {Klein}, {Koirala}, {Kolanoski}, {K{\"o}pke}, {Kopper}, {Kopper},
  {Koschinsky}, {Koskinen}, {Kowalski}, {Krings}, {Kroll}, {Kr{\"u}ckl},
  {Kunwar}, {Kurahashi}, {Kuwabara}, {Kyriacou}, {Labare}, {Lanfranchi},
  {Larson}, {Lauber}, {Leonard}, {Lesiak-Bzdak}, {Leuermann}, {Liu}, {Lozano
  Mariscal}, {Lu}, {L{\"u}nemann}, {Luszczak}, {Madsen}, {Maggi}, {Mahn},
  {Mancina}, {Maruyama}, {Mase}, {Maunu}, {Meagher}, {Medici}, {Meier},
  {Menne}, {Merino}, {Meures}, {Miarecki}, {Micallef}, {Moment{\'e}},
  {Montaruli}, {Moore}, {S}, {Morse}, {Moulai}, {Nahnhauer}, {Nakarmi},
  {Naumann}, {Neer}, {Niederhausen}, {Nowicki}, {Nygren}, {Obertacke Pollmann},
  {Olivas}, {O'Murchadha}, {O'Sullivan}, {Palczewski}, {Pandya}, {Pankova},
  {Peiffer}, {Pepper}, {P{\'e}rez de los Heros}, {Pieloth}, {Pinat}, {Plum},
  {Price}, {Przybylski}, {Raab}, {R{\"a}del}, {Rameez}, {Rauch}, {Rawlins},
  {Rea}, {Reimann}, {Relethford}, {Relich}, {Resconi}, {Rhode}, {Richman},
  {Robertson}, {Rongen}, {Rott}, {Ruhe}, {Ryckbosch}, {Rysewyk}, {Safa},
  {S{\"a}lzer}, {Sanchez Herrera}, {Sandrock}, {Sandroos}, {Santander},
  {Sarkar}, {Sarkar}, {Satalecka}, {Schlunder}, {Schmidt}, {Schneider},
  {Schoenen}, {Sch{\"o}neberg}, {Schumacher}, {Sclafani}, {Seckel},
  {Seunarine}, {Soedingrekso}, {Soldin}, {Song}, {Spiczak}, {Spiering},
  {Stachurska}, {Stamatikos}, {Stanev}, {Stasik}, {Stein}, {Stettner},
  {Steuer}, {Stezelberger}, {Stokstad}, {St{\"o}{\ss}l}, {Strotjohann},
  {Stuttard}, {Sullivan}, {Sutherland}, {Taboada}, {Tatar}, {Tenholt},
  {Ter-Antonyan}, {Terliuk}, {Tilav}, {Toale}, {Tobin}, {Toennis}, {Toscano},
  {Tosi}, {Tselengidou}, {Tung}, {Turcati}, {Turley}, {Ty}, {Unger}, {Usner},
  {Vandenbroucke}, {Van Driessche}, {van Eijk}, {van Eijndhoven}, {Vanheule},
  {van Santen}, {Vogel}, {Vraeghe}, {Walck}, {Wallace}, {Wallraff}, {Wandler},
  {Wandkowsky}, {Waza}, {Weaver}, {Weiss}, {Wendt}, {Werthebach}, {Westerhoff},
  {Whelan}, {Whitehorn}, {Wiebe}, {Wiebusch}, {Wille}, {Williams}, {Wills},
  {Wolf}, {Wood}, {Wood}, {Woschnagg}, {Xu}, {Xu}, {Xu}, {Yanez}, {Yodh},
  {Yoshida}, {Yuan}, {Fermi-LAT Collaboration}, {Abdollahi}, {Ajello},
  {Angioni}, {Baldini}, {Ballet}, {Barbiellini}, {Bastieri}, {Bechtol},
  {Bellazzini}, {Berenji}, {Bissaldi}, {Blandford}, {Bonino}, {Bottacini},
  {Bregeon}, {Bruel}, {Buehler}, {Burnett}, {Burns}, {Buson}, {Cameron},
  {Caputo}, {Caraveo}, {Cavazzuti}, {Charles}, {Chen}, {Cheung}, {Chiang},
  {Chiaro}, {Ciprini}, {Cohen-Tanugi}, {Conrad}, {Costantin}, {Cutini},
  {D'Ammando}, {de Palma}, {Digel}, {Di Lalla}, {Di Mauro}, {Di Venere},
  {Dom{\'\i}nguez}, {Favuzzi}, {Franckowiak}, {Fukazawa}, {Funk}, {Fusco},
  {Gargano}, {Gasparrini}, {Giglietto}, {Giomi}, {Giommi}, {Giordano},
  {Giroletti}, {Glanzman}, {Green}, {Grenier}, {Grondin}, {Guiriec}, {Harding},
  {Hayashida}, {Hays}, {Hewitt}, {Horan}, {J{\'o}hannesson}, {Kadler},
  {Kensei}, {Kocevski}, {Krauss}, {Kreter}, {Kuss}, {La Mura}, {Larsson},
  {Latronico}, {Lemoine-Goumard}, {Li}, {Longo}, {Loparco}, {Lovellette},
  {Lubrano}, {Magill}, {Maldera}, {Malyshev}, {Manfreda}, {Mazziotta},
  {McEnery}, {Meyer}, {Michelson}, {Mizuno}, {Monzani}, {Morselli},
  {Moskalenko}, {Negro}, {Nuss}, {Ojha}, {Omodei}, {Orienti}, {Orlando},
  {Palatiello}, {Paliya}, {Perkins}, {Persic}, {Pesce-Rollins}, {Piron},
  {Porter}, {Principe}, {Rain{\`o}}, {Rando}, {Rani}, {Razzano}, {Razzaque},
  {Reimer}, {Reimer}, {Renault-Tinacci}, {Ritz}, {Rochester}, {Saz Parkinson},
  {Sgr{\`o}}, {Siskind}, {Spandre}, {Spinelli}, {Suson}, {Tajima}, {Takahashi},
  {Tanaka}, {Thayer}, {Thompson}, {Tibaldo}, {Torres}, {Torresi}, {Tosti},
  {Troja}, {Valverde}, {Vianello}, {Vogel}, {Wood}, {Wood}, {Zaharijas}, {MAGIC
  Collaboration}, {Ahnen}, {Ansoldi}, {Antonelli}, {Arcaro}, {Baack},
  {Babi{\'c}}, {Banerjee}, {Bangale}, {Barres de Almeida}, {Barrio}, {Becerra
  Gonz{\'a}lez}, {Bednarek}, {Bernardini}, {Berti}, {Bhattacharyya}, {Biland},
  {Blanch}, {Bonnoli}, {Carosi}, {Carosi}, {Ceribella}, {Chatterjee}, {Colak},
  {Colin}, {Colombo}, {Contreras}, {Cortina}, {Covino}, {Cumani}, {Da Vela},
  {Dazzi}, {De Angelis}, {De Lotto}, {Delfino}, {Delgado}, {Di Pierro},
  {Dom{\'\i}nguez}, {Dominis Prester}, {Dorner}, {Doro}, {Einecke},
  {Elsaesser}, {Fallah Ramazani}, {Fern{\'a}ndez-Barral}, {Fidalgo}, {Foffano},
  {Pfrang}, {Fonseca}, {Font}, {Franceschini}, {Fruck}, {Galindo}, {Gallozzi},
  {Garc{\'\i}a L{\'o}pez}, {Garczarczyk}, {Gaug}, {Giammaria}, {Godinovi{\'c}},
  {Gora}, {Guberman}, {Hadasch}, {Hahn}, {Hassan}, {Hayashida}, {Herrera},
  {Hose}, {Hrupec}, {Inoue}, {Ishio}, {Konno}, {Kubo}, {Kushida}, {Lelas},
  {Lindfors}, {Lombardi}, {Longo}, {L{\'o}pez}, {Maggio}, {Majumdar},
  {Makariev}, {Maneva}, {Manganaro}, {Mannheim}, {Maraschi}, {Mariotti},
  {Mart{\'\i}nez}, {Masuda}, {Mazin}, {Minev}, {M}, {Mirzoyan}, {Moralejo},
  {Moreno}, {Moretti}, {Nagayoshi}, {Neustroev}, {Niedzwiecki}, {Nievas
  Rosillo}, {Nigro}, {Nilsson}, {Ninci}, {Nishijima}, {Noda}, {Nogu{\'e}s},
  {Paiano}, {Palacio}, {Paneque}, {Paoletti}, {Paredes}, {Pedaletti},
  {Peresano}, {Persic}, {Prada Moroni}, {Prandini}, {Puljak}, {Rodriguez
  Garcia}, {Reichardt}, {Rhode}, {Rib{\'o}}, {Rico}, {Righi}, {Rugliancich},
  {Saito}, {Satalecka}, {Schweizer}, {Sitarek}, {{\v{S}}nidaric ́},
  {Sobczynska}, {Stamerra}, {Strzys}, {Suri{\'c}}, {Takahashi}, {Tavecchio},
  {Temnikov}, {Terzi{\'c}}, {Teshima}, {Torres-Alb{\`a}}, {Treves},
  {Tsujimoto}, {Vanzo}, {Vazquez Acosta}, {Vovk}, {Ward}, {Will}, {S}, {Zaric
  ́}, {AGILE Team}, {Lucarelli}, {Tavani}, {Piano}, {Donnarumma}, {Pittori},
  {Verrecchia}, {Barbiellini}, {Bulgarelli}, {Caraveo}, {Cattaneo},
  {Colafrancesco}, {Costa}, {Di Cocco}, {Ferrari}, {Gianotti}, {Giuliani},
  {Lipari}, {Mereghetti}, {Morselli}, {Pacciani}, {Paoletti}, {Parmiggiani},
  {Pellizzoni}, {Picozza}, {Pilia}, {Rappoldi}, {Trois}, {Vercellone},
  {Vittorini}, {ASAS-SN Team}, {Stanek}, {Kochanek}, {Beacom}, {Thompson},
  {Holoien}, {Dong}, {Prieto}, {Shappee}, {Holmbo}, {HAWC Collaboration},
  {Abeysekara}, {Albert}, {Alfaro}, {Alvarez}, {Arceo},
  {Arteaga-Vel{\'a}zquez}, {Avila Rojas}, {Ayala Solares}, {Becerril},
  {Belmont-Moreno}, {Bernal}, {Caballero-Mora}, {Capistr{\'a}n},
  {Carrami{\~n}ana}, {Casanova}, {Castillo}, {Cotti}, {Cotzomi}, {Couti{\~n}o
  de Le{\'o}n}, {De Le{\'o}n}, {De la Fuente}, {Diaz Hernandez}, {Dichiara},
  {Dingus}, {DuVernois}, {D{\'\i}az-V{\'e}lez}, {Ellsworth}, {Engel},
  {Fiorino}, {Fleischhack}, {Fraija}, {Garc{\'\i}a-Gonz{\'a}lez}, {Garfias},
  {Gonz{\'a}lez Mu{\~n}oz}, {Gonz{\'a}lez}, {Goodman}, {Hampel-Arias},
  {Harding}, {Hernand ez}, {Hona}, {Hueyotl-Zahuantitla}, {Hui},
  {H{\"u}ntemeyer}, {Iriarte}, {Jardin-Blicq}, {Joshi}, {Kaufmann}, {Kunde},
  {Lara}, {Lauer}, {Lee}, {Lennarz}, {Le{\'o}n Vargas}, {Linnemann},
  {Longinotti}, {Luis-Raya}, {Luna-Garc{\'\i}a}, {Malone}, {Marinelli},
  {Martinez}, {Martinez-Castellanos}, {Mart{\'\i}nez-Castro},
  {Mart{\'\i}nez-Huerta}, {Matthews}, {Miranda-Romagnoli}, {Moreno},
  {Mostaf{\'a}}, {Nayerhoda}, {Nellen}, {Newbold}, {Nisa}, {Noriega-Papaqui},
  {Pelayo}, {Pretz}, {P{\'e}rez-P{\'e}rez}, {Ren}, {Rho}, {Rivi{\`e}re},
  {Rosa-Gonz{\'a}lez}, {Rosenberg}, {Ruiz-Velasco}, {Ruiz-Velasco}, {Salesa
  Greus}, {Sandoval}, {Schneider}, {Schoorlemmer}, {Sinnis}, {Smith},
  {Springer}, {Surajbali}, {Tibolla}, {Tollefson}, {Torres}, {Villase{\~n}or},
  {Weisgarber}, {Werner}, {Yapici}, {Gaurang}, {Zepeda}, {Zhou}, {{\'A}lvarez},
  {H.~E.~S.~S. Collaboration}, {Abdalla}, {Ang{\"u}ner}, {Armand}, {Backes},
  {Becherini}, {Berge}, {B{\"o}ttcher}, {Boisson}, {Bolmont}, {Bonnefoy},
  {Bordas}, {Brun}, {B{\"u}chele}, {Bulik}, {Caroff}, {Carosi}, {Casanova},
  {Cerruti}, {Chakraborty}, {Chandra}, {Chen}, {Colafrancesco}, {Davids},
  {Deil}, {Devin}, {Djannati-Ata{\"\i}}, {Egberts}, {Emery}, {Eschbach},
  {Fiasson}, {Fontaine}, {Funk}, {F{\"u}{\ss}ling}, {Gallant}, {Gat{\'e}},
  {Giavitto}, {Glawion}, {Glicenstein}, {Gottschall}, {Grondin}, {Haupt},
  {Henri}, {Hinton}, {Hoischen}, {Holch}, {Huber}, {Jamrozy}, {Jankowsky},
  {Jankowsky}, {Jouvin}, {Jung-Richardt}, {Kerszberg}, {Kh{\'e}lifi}, {King},
  {Klepser}, {Kluz ́niak}, {Komin}, {Kraus}, {Lefaucheur}, {Lemi{\`e}re},
  {Lemoine-Goumard}, {Lenain}, {Leser}, {Lohse}, {L{\'o}pez-Coto}, {Lorentz},
  {Lypova}, {Marandon}, {Guillem Mart{\'\i}-Devesa}, {Maurin}, {Mitchell},
  {Moderski}, {Mohamed}, {Mohrmann}, {Moulin}, {Murach}, {de Naurois},
  {Niederwanger}, {Niemiec}, {Oakes}, {O'Brien}, {Ohm}, {Ostrowski}, {Oya},
  {Panter}, {Parsons}, {Perennes}, {Piel}, {Pita}, {Poireau}, {Priyana Noel},
  {Prokoph}, {P{\"u}hlhofer}, {Quirrenbach}, {Raab}, {Rauth}, {Renaud},
  {Rieger}, {Rinchiuso}, {Romoli}, {Rowell}, {Rudak}, {Sasaki}, {Sanchez},
  {Schlickeiser}, {Sch{\"u}ssler}, {Schulz}, {Schwanke}, {Seglar-Arroyo},
  {Shafi}, {Simoni}, {Sol}, {Stegmann}, {Steppa}, {Tavernier}, {Taylor},
  {Tiziani}, {Trichard}, {Tsirou}, {van Eldik}, {van Rensburg}, {van Soelen},
  {Veh}, {Vincent}, {Voisin}, {Wagner}, {Wagner}, {Wierzcholska}, {Zanin},
  {Zdziarski}, {Zech}, {Ziegler}, {Zorn}, {{\.Z}ywucka}, {INTEGRAL Team},
  {Savchenko}, {Ferrigno}, {Bazzano}, {Diehl}, {Kuulkers}, {Laurent},
  {Mereghetti}, {Natalucci}, {Panessa}, {Rodi}, {Ubertini}, {Kanata}, Teams,
  {Morokuma}, {Ohta}, {Tanaka}, {Mori}, {Yamanaka}, {Kawabata}, {Utsumi},
  {Nakaoka}, {Kawabata}, {Nagashima}, {Yoshida}, {Matsuoka}, {Itoh}, {Kapteyn
  Team}, {Keel}, {Liverpool Telescope Team}, {Copperwheat}, {Steele},
  {Swift/NuSTAR Team}, {Cenko}, {Cowen}, {DeLaunay}, {Evans}, {Fox}, {Keivani},
  {Kennea}, {Marshall}, {Osborne}, {Santander}, {Tohuvavohu}, {Turley},
  {VERITAS Collaboration}, {Abeysekara}, {Archer}, {Benbow}, {Bird}, {Brill},
  {Brose}, {Buchovecky}, {Buckley}, {Bugaev}, {Christiansen}, {Connolly},
  {Cui}, {Daniel}, {Errando}, {Falcone}, {Feng}, {Finley}, {Fortson},
  {Furniss}, {Gueta}, {H{\"u}tten}, {Hervet}, {Hughes}, {Humensky}, {Johnson},
  {Kaaret}, {Kar}, {Kelley-Hoskins}, {Kertzman}, {Kieda}, {Krause},
  {Krennrich}, {Kumar}, {Lang}, {Lin}, {Maier}, {McArthur}, {Moriarty},
  {Mukherjee}, {Nieto}, {O'Brien}, {Ong}, {Otte}, {Park}, {Petrashyk}, {Pohl},
  {Popkow}, {Pueschel}, {Quinn}, {Ragan}, {Reynolds}, {Richards}, {Roache},
  {Rulten}, {Sadeh}, {Santander}, {Scott}, {Sembroski}, {Shahinyan}, {Sushch},
  {Tr{\'e}panier}, {Tyler}, {Vassiliev}, {Wakely}, {Weinstein}, {Wells},
  {Wilcox}, {Wilhelm}, {Williams}, {Zitzer}, {VLA/B Team}, {Tetarenko},
  {Kimball}, {Miller-Jones}, \& {Sivakoff}}]{2018Sci...361.1378I}
{IceCube Collaboration}, {Aartsen}, M.~G., {Ackermann}, M., {et~al.} 2018,
  Science, 361, eaat1378, \dodoi{10.1126/science.aat1378}

\bibitem[{{Ivezi{\'c}} {et~al.}(2019){Ivezi{\'c}}, {Kahn}, {Tyson}, {Abel},
  {Acosta}, {Allsman}, {Alonso}, {AlSayyad}, {Anderson}, {Andrew}, {Angel},
  {Angeli}, {Ansari}, {Antilogus}, {Araujo}, {Armstrong}, {Arndt}, {Astier},
  {Aubourg}, {Auza}, {Axelrod}, {Bard}, {Barr}, {Barrau}, {Bartlett}, {Bauer},
  {Bauman}, {Baumont}, {Bechtol}, {Bechtol}, {Becker}, {Becla}, {Beldica},
  {Bellavia}, {Bianco}, {Biswas}, {Blanc}, {Blazek}, {Bland ford}, {Bloom},
  {Bogart}, {Bond}, {Booth}, {Borgland}, {Borne}, {Bosch}, {Boutigny},
  {Brackett}, {Bradshaw}, {Brand t}, {Brown}, {Bullock}, {Burchat}, {Burke},
  {Cagnoli}, {Calabrese}, {Callahan}, {Callen}, {Carlin}, {Carlson}, {Chand
  rasekharan}, {Charles-Emerson}, {Chesley}, {Cheu}, {Chiang}, {Chiang},
  {Chirino}, {Chow}, {Ciardi}, {Claver}, {Cohen-Tanugi}, {Cockrum}, {Coles},
  {Connolly}, {Cook}, {Cooray}, {Covey}, {Cribbs}, {Cui}, {Cutri}, {Daly},
  {Daniel}, {Daruich}, {Daubard}, {Daues}, {Dawson}, {Delgado}, {Dellapenna},
  {de Peyster}, {de Val-Borro}, {Digel}, {Doherty}, {Dubois},
  {Dubois-Felsmann}, {Durech}, {Economou}, {Eifler}, {Eracleous}, {Emmons},
  {Fausti Neto}, {Ferguson}, {Figueroa}, {Fisher-Levine}, {Focke}, {Foss},
  {Frank}, {Freemon}, {Gangler}, {Gawiser}, {Geary}, {Gee}, {Geha}, {Gessner},
  {Gibson}, {Gilmore}, {Glanzman}, {Glick}, {Goldina}, {Goldstein}, {Goodenow},
  {Graham}, {Gressler}, {Gris}, {Guy}, {Guyonnet}, {Haller}, {Harris},
  {Hascall}, {Haupt}, {Hernand ez}, {Herrmann}, {Hileman}, {Hoblitt},
  {Hodgson}, {Hogan}, {Howard}, {Huang}, {Huffer}, {Ingraham}, {Innes},
  {Jacoby}, {Jain}, {Jammes}, {Jee}, {Jenness}, {Jernigan}, {Jevremovi{\'c}},
  {Johns}, {Johnson}, {Johnson}, {Jones}, {Juramy-Gilles}, {Juri{\'c}},
  {Kalirai}, {Kallivayalil}, {Kalmbach}, {Kantor}, {Karst}, {Kasliwal},
  {Kelly}, {Kessler}, {Kinnison}, {Kirkby}, {Knox}, {Kotov}, {Krabbendam},
  {Krughoff}, {Kub{\'a}nek}, {Kuczewski}, {Kulkarni}, {Ku}, {Kurita}, {Lage},
  {Lambert}, {Lange}, {Langton}, {Le Guillou}, {Levine}, {Liang}, {Lim},
  {Lintott}, {Long}, {Lopez}, {Lotz}, {Lupton}, {Lust}, {MacArthur}, {Mahabal},
  {Mand elbaum}, {Markiewicz}, {Marsh}, {Marshall}, {Marshall}, {May},
  {McKercher}, {McQueen}, {Meyers}, {Migliore}, {Miller}, {Mills}, {Miraval},
  {Moeyens}, {Moolekamp}, {Monet}, {Moniez}, {Monkewitz}, {Montgomery},
  {Morrison}, {Mueller}, {Muller}, {Mu{\~n}oz Arancibia}, {Neill}, {Newbry},
  {Nief}, {Nomerotski}, {Nordby}, {O'Connor}, {Oliver}, {Olivier}, {Olsen},
  {O'Mullane}, {Ortiz}, {Osier}, {Owen}, {Pain}, {Palecek}, {Parejko},
  {Parsons}, {Pease}, {Peterson}, {Peterson}, {Petravick}, {Libby Petrick},
  {Petry}, {Pierfederici}, {Pietrowicz}, {Pike}, {Pinto}, {Plante}, {Plate},
  {Plutchak}, {Price}, {Prouza}, {Radeka}, {Rajagopal}, {Rasmussen},
  {Regnault}, {Reil}, {Reiss}, {Reuter}, {Ridgway}, {Riot}, {Ritz}, {Robinson},
  {Roby}, {Roodman}, {Rosing}, {Roucelle}, {Rumore}, {Russo}, {Saha},
  {Sassolas}, {Schalk}, {Schellart}, {Schindler}, {Schmidt}, {Schneider},
  {Schneider}, {Schoening}, {Schumacher}, {Schwamb}, {Sebag}, {Selvy},
  {Sembroski}, {Seppala}, {Serio}, {Serrano}, {Shaw}, {Shipsey}, {Sick},
  {Silvestri}, {Slater}, {Smith}, {Smith}, {Sobhani}, {Soldahl},
  {Storrie-Lombardi}, {Stover}, {Strauss}, {Street}, {Stubbs}, {Sullivan},
  {Sweeney}, {Swinbank}, {Szalay}, {Takacs}, {Tether}, {Thaler}, {Thayer},
  {Thomas}, {Thornton}, {Thukral}, {Tice}, {Trilling}, {Turri}, {Van Berg},
  {Vanden Berk}, {Vetter}, {Virieux}, {Vucina}, {Wahl}, {Walkowicz}, {Walsh},
  {Walter}, {Wang}, {Wang}, {Warner}, {Wiecha}, {Willman}, {Winters},
  {Wittman}, {Wolff}, {Wood-Vasey}, {Wu}, {Xin}, {Yoachim}, \&
  {Zhan}}]{2019ApJ...873..111I}
{Ivezi{\'c}}, {\v{Z}}., {Kahn}, S.~M., {Tyson}, J.~A., {et~al.} 2019, \apj,
  873, 111, \dodoi{10.3847/1538-4357/ab042c}

\bibitem[{{Kellermann} {et~al.}(1989){Kellermann}, {Sramek}, {Schmidt},
  {Shaffer}, \& {Green}}]{1989AJ.....98.1195K}
{Kellermann}, K.~I., {Sramek}, R., {Schmidt}, M., {Shaffer}, D.~B., \& {Green},
  R. 1989, \aj, 98, 1195, \dodoi{10.1086/115207}

\bibitem[{{Kilerci Eser} {et~al.}(2015){Kilerci Eser}, {Vestergaard},
  {Peterson}, {Denney}, \& {Bentz}}]{2015ApJ...801....8K}
{Kilerci Eser}, E., {Vestergaard}, M., {Peterson}, B.~M., {Denney}, K.~D., \&
  {Bentz}, M.~C. 2015, \apj, 801, 8, \dodoi{10.1088/0004-637X/801/1/8}

\bibitem[{{Li} {et~al.}(2018){Li}, {Xia}, {Liang}, {Liao}, \&
  {Fan}}]{2018ApJ...853..159L}
{Li}, S., {Xia}, Z.-Q., {Liang}, Y.-F., {Liao}, N.-H., \& {Fan}, Y.-Z. 2018,
  \apj, 853, 159, \dodoi{10.3847/1538-4357/aaa3fb}

\bibitem[{{Liao} {et~al.}(2014){Liao}, {Bai}, {Liu}, {Weng}, {Chen}, \&
  {Li}}]{2014ApJ...783...83L}
{Liao}, N.~H., {Bai}, J.~M., {Liu}, H.~T., {et~al.} 2014, \apj, 783, 83,
  \dodoi{10.1088/0004-637X/783/2/83}

\bibitem[{{Liao} {et~al.}(2019){Liao}, {Dou}, {Jiang}, {Wang}, {Fan}, \&
  {Wang}}]{2019ApJ...879L...9L}
{Liao}, N.-H., {Dou}, L.-M., {Jiang}, N., {et~al.} 2019, \apjl, 879, L9,
  \dodoi{10.3847/2041-8213/ab2893}

\bibitem[{{Liao} {et~al.}(2018){Liao}, {Li}, \& {Fan}}]{2018ApJ...865L..17L}
{Liao}, N.-H., {Li}, S., \& {Fan}, Y.-Z. 2018, \apjl, 865, L17,
  \dodoi{10.3847/2041-8213/aae20d}

\bibitem[{{Liao} {et~al.}(2016){Liao}, {Xin}, {Fan}, {Weng}, {Li}, {Chen}, \&
  {Fan}}]{2016ApJS..226...17L}
{Liao}, N.-H., {Xin}, Y.-L., {Fan}, X.-L., {et~al.} 2016, \apjs, 226, 17,
  \dodoi{10.3847/0067-0049/226/2/17}

\bibitem[{{Lou} {et~al.}(2016){Lou}, {Liang}, {Yao}, {Zheng}, {Cheng}, {Wang},
  {Liu}, {Qian}, {Zhao}, \& {Yang}}]{2016SPIE10154E..2AL}
{Lou}, Z., {Liang}, M., {Yao}, D., {et~al.} 2016, in Society of Photo-Optical
  Instrumentation Engineers (SPIE) Conference Series, Vol. 10154, \procspie,
  101542A, \dodoi{10.1117/12.2248371}

\bibitem[{{Madejski} \& {Sikora}(2016)}]{2016ARA&A..54..725M}
{Madejski}, G.~G., \& {Sikora}, M. 2016, \araa, 54, 725,
  \dodoi{10.1146/annurev-astro-081913-040044}

\bibitem[{{Maraschi} {et~al.}(1992){Maraschi}, {Ghisellini}, \&
  {Celotti}}]{1992ApJ...397L...5M}
{Maraschi}, L., {Ghisellini}, G., \& {Celotti}, A. 1992, \apjl, 397, L5,
  \dodoi{10.1086/186531}

\bibitem[{{Marcotulli} {et~al.}(2020){Marcotulli}, {Paliya}, {Ajello}, {Kaur},
  {Marchesi}, {Rajagopal}, {Hartmann}, {Gasparrini}, {Ojha}, \&
  {Madejski}}]{2020ApJ...889..164M}
{Marcotulli}, L., {Paliya}, V., {Ajello}, M., {et~al.} 2020, \apj, 889, 164,
  \dodoi{10.3847/1538-4357/ab65f5}

\bibitem[{{Masci} {et~al.}(2019){Masci}, {Laher}, {Rusholme}, {Shupe}, {Groom},
  {Surace}, {Jackson}, {Monkewitz}, {Beck}, {Flynn}, {Terek}, {Landry},
  {Hacopians}, {Desai}, {Howell}, {Brooke}, {Imel}, {Wachter}, {Ye}, {Lin},
  {Cenko}, {Cunningham}, {Rebbapragada}, {Bue}, {Miller}, {Mahabal}, {Bellm},
  {Patterson}, {Juri{\'c}}, {Golkhou}, {Ofek}, {Walters}, {Graham}, {Kasliwal},
  {Dekany}, {Kupfer}, {Burdge}, {Cannella}, {Barlow}, {Van Sistine}, {Giomi},
  {Fremling}, {Blagorodnova}, {Levitan}, {Riddle}, {Smith}, {Helou}, {Prince},
  \& {Kulkarni}}]{2019PASP..131a8003M}
{Masci}, F.~J., {Laher}, R.~R., {Rusholme}, B., {et~al.} 2019, \pasp, 131,
  018003, \dodoi{10.1088/1538-3873/aae8ac}

\bibitem[{{Massaro} {et~al.}(2009){Massaro}, {Giommi}, {Leto}, {Marchegiani},
  {Maselli}, {Perri}, {Piranomonte}, \& {Sclavi}}]{2009A&A...495..691M}
{Massaro}, E., {Giommi}, P., {Leto}, C., {et~al.} 2009, \aap, 495, 691,
  \dodoi{10.1051/0004-6361:200810161}

\bibitem[{{Mathur} \& {Elvis}(1995)}]{1995AJ....110.1551M}
{Mathur}, S., \& {Elvis}, M. 1995, \aj, 110, 1551, \dodoi{10.1086/117627}

\bibitem[{{Mattox} {et~al.}(1996){Mattox}, {Bertsch}, {Chiang}, {Dingus},
  {Digel}, {Esposito}, {Fierro}, {Hartman}, {Hunter}, {Kanbach}, {Kniffen},
  {Lin}, {Macomb}, {Mayer-Hasselwander}, {Michelson}, {von Montigny},
  {Mukherjee}, {Nolan}, {Ramanamurthy}, {Schneid}, {Sreekumar}, {Thompson}, \&
  {Willis}}]{1996ApJ...461..396M}
{Mattox}, J.~R., {Bertsch}, D.~L., {Chiang}, J., {et~al.} 1996, \apj, 461, 396,
  \dodoi{10.1086/177068}

\bibitem[{{McEnery} \& {Amego Team}(2020)}]{2020AAS...23537215M}
{McEnery}, J., \& {Amego Team}. 2020, in American Astronomical Society Meeting
  Abstracts, American Astronomical Society Meeting Abstracts, 372.15

\bibitem[{{Moran} \& {Helfand}(1997)}]{1997ApJ...484L..95M}
{Moran}, E.~C., \& {Helfand}, D.~J. 1997, \apjl, 484, L95,
  \dodoi{10.1086/310787}

\bibitem[{{Nolan} {et~al.}(2012){Nolan}, {Abdo}, {Ackermann}, {Ajello},
  {Allafort}, {Antolini}, {Atwood}, {Axelsson}, {Baldini}, {Ballet},
  {Barbiellini}, {Bastieri}, {Bechtol}, {Belfiore}, {Bellazzini}, {Berenji},
  {Bignami}, {Blandford}, {Bloom}, {Bonamente}, {Bonnell}, {Borgland},
  {Bottacini}, {Bouvier}, {Brandt}, {Bregeon}, {Brigida}, {Bruel}, {Buehler},
  {Burnett}, {Buson}, {Caliandro}, {Cameron}, {Campana}, {Ca{\~n}adas},
  {Cannon}, {Caraveo}, {Casandjian}, {Cavazzuti}, {Ceccanti}, {Cecchi},
  {{\c{C}}elik}, {Charles}, {Chekhtman}, {Cheung}, {Chiang}, {Chipaux},
  {Ciprini}, {Claus}, {Cohen-Tanugi}, {Cominsky}, {Conrad}, {Corbet}, {Cutini},
  {D'Ammando}, {Davis}, {de Angelis}, {DeCesar}, {DeKlotz}, {De Luca}, {den
  Hartog}, {de Palma}, {Dermer}, {Digel}, {Silva}, {Drell}, {Drlica-Wagner},
  {Dubois}, {Dumora}, {Enoto}, {Escande}, {Fabiani}, {Falletti}, {Favuzzi},
  {Fegan}, {Ferrara}, {Focke}, {Fortin}, {Frailis}, {Fukazawa}, {Funk},
  {Fusco}, {Gargano}, {Gasparrini}, {Gehrels}, {Germani}, {Giebels},
  {Giglietto}, {Giommi}, {Giordano}, {Giroletti}, {Glanzman}, {Godfrey},
  {Grenier}, {Grondin}, {Grove}, {Guillemot}, {Guiriec}, {Gustafsson},
  {Hadasch}, {Hanabata}, {Harding}, {Hayashida}, {Hays}, {Hill}, {Horan},
  {Hou}, {Hughes}, {Iafrate}, {Itoh}, {J{\'o}hannesson}, {Johnson}, {Johnson},
  {Johnson}, {Johnson}, {Kamae}, {Katagiri}, {Kataoka}, {Katsuta}, {Kawai},
  {Kerr}, {Kn{\"o}dlseder}, {Kocevski}, {Kuss}, {Lande}, {Landriu},
  {Latronico}, {Lemoine-Goumard}, {Lionetto}, {Llena Garde}, {Longo},
  {Loparco}, {Lott}, {Lovellette}, {Lubrano}, {Madejski}, {Marelli}, {Massaro},
  {Mazziotta}, {McConville}, {McEnery}, {Mehault}, {Michelson}, {Minuti},
  {Mitthumsiri}, {Mizuno}, {Moiseev}, {Mongelli}, {Monte}, {Monzani},
  {Morselli}, {Moskalenko}, {Murgia}, {Nakamori}, {Naumann-Godo}, {Norris},
  {Nuss}, {Nymark}, {Ohno}, {Ohsugi}, {Okumura}, {Omodei}, {Orlando}, {Ormes},
  {Ozaki}, {Paneque}, {Panetta}, {Parent}, {Perkins}, {Pesce-Rollins},
  {Pierbattista}, {Pinchera}, {Piron}, {Pivato}, {Porter}, {Racusin},
  {Rain{\`o}}, {Rand o}, {Razzano}, {Razzaque}, {Reimer}, {Reimer}, {Reposeur},
  {Ritz}, {Rochester}, {Romani}, {Roth}, {Rousseau}, {Ryde}, {Sadrozinski},
  {Salvetti}, {Sanchez}, {Saz Parkinson}, {Sbarra}, {Scargle}, {Schalk},
  {Sgr{\`o}}, {Shaw}, {Shrader}, {Siskind}, {Smith}, {Spandre}, {Spinelli},
  {Stephens}, {Strickman}, {Suson}, {Tajima}, {Takahashi}, {Takahashi},
  {Tanaka}, {Thayer}, {Thayer}, {Thompson}, {Tibaldo}, {Tibolla}, {Tinebra},
  {Tinivella}, {Torres}, {Tosti}, {Troja}, {Uchiyama}, {Vandenbroucke}, {Van
  Etten}, {Van Klaveren}, {Vasileiou}, {Vianello}, {Vitale}, {Waite},
  {Wallace}, {Wang}, {Werner}, {Winer}, {Wood}, {Wood}, {Wood}, {Yang}, \&
  {Zimmer}}]{2012ApJS..199...31N}
{Nolan}, P.~L., {Abdo}, A.~A., {Ackermann}, M., {et~al.} 2012, \apjs, 199, 31,
  \dodoi{10.1088/0067-0049/199/2/31}

\bibitem[{{Ofek} {et~al.}(2012){Ofek}, {Laher}, {Surace}, {Levitan}, {Sesar},
  {Horesh}, {Law}, {van Eyken}, {Kulkarni}, {Prince}, {Nugent}, {Sullivan},
  {Yaron}, {Pickles}, {Ag{\"u}eros}, {Arcavi}, {Bildsten}, {Bloom}, {Cenko},
  {Gal-Yam}, {Grillmair}, {Helou}, {Kasliwal}, {Poznanski}, \&
  {Quimby}}]{2012PASP..124..854O}
{Ofek}, E.~O., {Laher}, R., {Surace}, J., {et~al.} 2012, \pasp, 124, 854,
  \dodoi{10.1086/666978}

\bibitem[{{Oh} {et~al.}(2018){Oh}, {Koss}, {Markwardt}, {Schawinski},
  {Baumgartner}, {Barthelmy}, {Cenko}, {Gehrels}, {Mushotzky}, {Petulante},
  {Ricci}, {Lien}, \& {Trakhtenbrot}}]{2018ApJS..235....4O}
{Oh}, K., {Koss}, M., {Markwardt}, C.~B., {et~al.} 2018, \apjs, 235, 4,
  \dodoi{10.3847/1538-4365/aaa7fd}

\bibitem[{{P{\^a}ris} {et~al.}(2014){P{\^a}ris}, {Petitjean}, {Aubourg},
  {Ross}, {Myers}, {Streblyanska}, {Bailey}, {Hall}, {Strauss}, {Anderson},
  {Bizyaev}, {Borde}, {Brinkmann}, {Bovy}, {Brandt}, {Brewington},
  {Brownstein}, {Cook}, {Ebelke}, {Fan}, {Filiz Ak}, {Finley}, {Font-Ribera},
  {Ge}, {Hamann}, {Ho}, {Jiang}, {Kinemuchi}, {Malanushenko}, {Malanushenko},
  {Marchante}, {McGreer}, {McMahon}, {Miralda-Escud{\'e}}, {Muna},
  {Noterdaeme}, {Oravetz}, {Palanque-Delabrouille}, {Pan}, {Perez-Fournon},
  {Pieri}, {Riffel}, {Schlegel}, {Schneider}, {Simmons}, {Viel}, {Weaver},
  {Wood-Vasey}, {Y{\`e}che}, \& {York}}]{2014A&A...563A..54P}
{P{\^a}ris}, I., {Petitjean}, P., {Aubourg}, {\'E}., {et~al.} 2014, \aap, 563,
  A54, \dodoi{10.1051/0004-6361/201322691}

\bibitem[{{Patnaik} {et~al.}(1992){Patnaik}, {Browne}, {Wilkinson}, \&
  {Wrobel}}]{1992MNRAS.254..655P}
{Patnaik}, A.~R., {Browne}, I. W.~A., {Wilkinson}, P.~N., \& {Wrobel}, J.~M.
  1992, \mnras, 254, 655, \dodoi{10.1093/mnras/254.4.655}

\bibitem[{{Planck Collaboration} {et~al.}(2014){Planck Collaboration}, {Ade},
  {Aghanim}, {Armitage-Caplan}, {Arnaud}, {Ashdown}, {Atrio-Barand ela},
  {Aumont}, {Baccigalupi}, {Banday}, {Barreiro}, {Bartlett}, {Battaner},
  {Benabed}, {Beno{\^\i}t}, {Benoit-L{\'e}vy}, {Bernard}, {Bersanelli},
  {Bielewicz}, {Bobin}, {Bock}, {Bonaldi}, {Bond}, {Borrill}, {Bouchet},
  {Bridges}, {Bucher}, {Burigana}, {Butler}, {Calabrese}, {Cappellini},
  {Cardoso}, {Catalano}, {Challinor}, {Chamballu}, {Chary}, {Chen}, {Chiang},
  {Chiang}, {Christensen}, {Church}, {Clements}, {Colombi}, {Colombo},
  {Couchot}, {Coulais}, {Crill}, {Curto}, {Cuttaia}, {Danese}, {Davies},
  {Davis}, {de Bernardis}, {de Rosa}, {de Zotti}, {Delabrouille}, {Delouis},
  {D{\'e}sert}, {Dickinson}, {Diego}, {Dolag}, {Dole}, {Donzelli}, {Dor{\'e}},
  {Douspis}, {Dunkley}, {Dupac}, {Efstathiou}, {Elsner}, {En{\ss}lin},
  {Eriksen}, {Finelli}, {Forni}, {Frailis}, {Fraisse}, {Franceschi}, {Gaier},
  {Galeotta}, {Galli}, {Ganga}, {Giard}, {Giardino}, {Giraud-H{\'e}raud},
  {Gjerl{\o}w}, {Gonz{\'a}lez-Nuevo}, {G{\'o}rski}, {Gratton}, {Gregorio},
  {Gruppuso}, {Gudmundsson}, {Haissinski}, {Hamann}, {Hansen}, {Hanson},
  {Harrison}, {Henrot-Versill{\'e}}, {Hern{\'a}ndez-Monteagudo}, {Herranz},
  {Hildebrand t}, {Hivon}, {Hobson}, {Holmes}, {Hornstrup}, {Hou}, {Hovest},
  {Huffenberger}, {Jaffe}, {Jaffe}, {Jewell}, {Jones}, {Juvela},
  {Keih{\"a}nen}, {Keskitalo}, {Kisner}, {Kneissl}, {Knoche}, {Knox}, {Kunz},
  {Kurki-Suonio}, {Lagache}, {L{\"a}hteenm{\"a}ki}, {Lamarre}, {Lasenby},
  {Lattanzi}, {Laureijs}, {Lawrence}, {Leach}, {Leahy}, {Leonardi},
  {Le{\'o}n-Tavares}, {Lesgourgues}, {Lewis}, {Liguori}, {Lilje},
  {Linden-V{\o}rnle}, {L{\'o}pez-Caniego}, {Lubin}, {Mac{\'\i}as-P{\'e}rez},
  {Maffei}, {Maino}, {Mand olesi}, {Maris}, {Marshall}, {Martin},
  {Mart{\'\i}nez-Gonz{\'a}lez}, {Masi}, {Massardi}, {Matarrese}, {Matthai},
  {Mazzotta}, {Meinhold}, {Melchiorri}, {Melin}, {Mendes}, {Menegoni},
  {Mennella}, {Migliaccio}, {Millea}, {Mitra}, {Miville-Desch{\^e}nes},
  {Moneti}, {Montier}, {Morgante}, {Mortlock}, {Moss}, {Munshi}, {Murphy},
  {Naselsky}, {Nati}, {Natoli}, {Netterfield}, {N{\o}rgaard-Nielsen},
  {Noviello}, {Novikov}, {Novikov}, {O'Dwyer}, {Osborne}, {Oxborrow}, {Paci},
  {Pagano}, {Pajot}, {Paladini}, {Paoletti}, {Partridge}, {Pasian},
  {Patanchon}, {Pearson}, {Pearson}, {Peiris}, {Perdereau}, {Perotto},
  {Perrotta}, {Pettorino}, {Piacentini}, {Piat}, {Pierpaoli}, {Pietrobon},
  {Plaszczynski}, {Platania}, {Pointecouteau}, {Polenta}, {Ponthieu}, {Popa},
  {Poutanen}, {Pratt}, {Pr{\'e}zeau}, {Prunet}, {Puget}, {Rachen}, {Reach},
  {Rebolo}, {Reinecke}, {Remazeilles}, {Renault}, {Ricciardi}, {Riller},
  {Ristorcelli}, {Rocha}, {Rosset}, {Roudier}, {Rowan-Robinson},
  {Rubi{\~n}o-Mart{\'\i}n}, {Rusholme}, {Sandri}, {Santos}, {Savelainen},
  {Savini}, {Scott}, {Seiffert}, {Shellard}, {Spencer}, {Starck}, {Stolyarov},
  {Stompor}, {Sudiwala}, {Sunyaev}, {Sureau}, {Sutton}, {Suur-Uski}, {Sygnet},
  {Tauber}, {Tavagnacco}, {Terenzi}, {Toffolatti}, {Tomasi}, {Tristram},
  {Tucci}, {Tuovinen}, {T{\"u}rler}, {Umana}, {Valenziano}, {Valiviita}, {Van
  Tent}, {Vielva}, {Villa}, {Vittorio}, {Wade}, {Wandelt}, {Wehus}, {White},
  {White}, {Wilkinson}, {Yvon}, {Zacchei}, \& {Zonca}}]{2014A&A...571A..16P}
{Planck Collaboration}, {Ade}, P.~A.~R., {Aghanim}, N., {et~al.} 2014, \aap,
  571, A16, \dodoi{10.1051/0004-6361/201321591}

\bibitem[{{Romani}(2006)}]{2006AJ....132.1959R}
{Romani}, R.~W. 2006, \aj, 132, 1959, \dodoi{10.1086/508216}

\bibitem[{{Romani} {et~al.}(2004){Romani}, {Sowards-Emmerd}, {Greenhill}, \&
  {Michelson}}]{2004ApJ...610L...9R}
{Romani}, R.~W., {Sowards-Emmerd}, D., {Greenhill}, L., \& {Michelson}, P.
  2004, \apjl, 610, L9, \dodoi{10.1086/423201}

\bibitem[{{Sbarrato} {et~al.}(2012){Sbarrato}, {Ghisellini}, {Nardini},
  {Tagliaferri}, {Foschini}, {Ghirlanda}, {Tavecchio}, {Greiner}, {Rau}, \&
  {Gehrels}}]{2012MNRAS.426L..91S}
{Sbarrato}, T., {Ghisellini}, G., {Nardini}, M., {et~al.} 2012, \mnras, 426,
  L91, \dodoi{10.1111/j.1745-3933.2012.01332.x}

\bibitem[{{Schinzel} {et~al.}(2011){Schinzel}, {Sokolovsky}, {D'Ammando},
  {Burnett}, {Max-Moerbeck}, {Cheung}, {Fegan}, {Casandjian}, {Reyes},
  {Villata}, {Raiteri}, {Agudo}, {Bravo Calle}, {Carosati}, {Casas},
  {G{\'o}mez}, {Gurwell}, {Hsiao}, {Jorstad}, {Kimeridze}, {Konstantinova},
  {Kopatskaya}, {Koptelova}, {Kurtanidze}, {Kurtanidze}, {Larionov},
  {Larionova}, {Larionova}, {Marscher}, {Morozova}, {Nikolashvili},
  {Roca-Sogorb}, {Ros}, {Sigua}, {Spiridonova}, {Troitsky}, {Vlasyuk},
  {Lobanov}, \& {Zensus}}]{2011A&A...532A.150S}
{Schinzel}, F.~K., {Sokolovsky}, K.~V., {D'Ammando}, F., {et~al.} 2011, \aap,
  532, A150, \dodoi{10.1051/0004-6361/201016145}

\bibitem[{{Schneider} {et~al.}(2005){Schneider}, {Hall}, {Richards}, {Vanden
  Berk}, {Anderson}, {Fan}, {Jester}, {Stoughton}, {Strauss}, {SubbaRao},
  {Brandt}, {Gunn}, {Yanny}, {Bahcall}, {Barentine}, {Blanton}, {Boroski},
  {Brewington}, {Brinkmann}, {Brunner}, {Csabai}, {Doi}, {Eisenstein},
  {Frieman}, {Fukugita}, {Gray}, {Harvanek}, {Heckman}, {Ivezi{\'c}}, {Kent},
  {Kleinman}, {Knapp}, {Kron}, {Krzesinski}, {Long}, {Loveday}, {Lupton},
  {Margon}, {Munn}, {Neilsen}, {Newberg}, {Newman}, {Nichol}, {Nitta}, {Pier},
  {Rockosi}, {Saxe}, {Schlegel}, {Snedden}, {Szalay}, {Thakar}, {Uomoto},
  {Voges}, \& {York}}]{2005AJ....130..367S}
{Schneider}, D.~P., {Hall}, P.~B., {Richards}, G.~T., {et~al.} 2005, \aj, 130,
  367, \dodoi{10.1086/431156}

\bibitem[{{Sikora} {et~al.}(1994){Sikora}, {Begelman}, \&
  {Rees}}]{1994ApJ...421..153S}
{Sikora}, M., {Begelman}, M.~C., \& {Rees}, M.~J. 1994, \apj, 421, 153,
  \dodoi{10.1086/173633}

\bibitem[{{Ulrich} {et~al.}(1997){Ulrich}, {Maraschi}, \&
  {Urry}}]{1997ARA&A..35..445U}
{Ulrich}, M.-H., {Maraschi}, L., \& {Urry}, C.~M. 1997, \araa, 35, 445,
  \dodoi{10.1146/annurev.astro.35.1.445}

\bibitem[{{Volonteri}(2010)}]{2010A&ARv..18..279V}
{Volonteri}, M. 2010, \aapr, 18, 279, \dodoi{10.1007/s00159-010-0029-x}

\bibitem[{{Yi} {et~al.}(2014){Yi}, {Wang}, {Wu}, {Yang}, {Bai}, {Fan},
  {Brandt}, {Ho}, {Zuo}, {Kim}, {Wang}, {Yang}, {Zhang}, {Wang}, {Wang}, {Ai},
  {Fan}, {Chang}, {Wang}, {Lun}, \& {Xin}}]{2014ApJ...795L..29Y}
{Yi}, W.-M., {Wang}, F., {Wu}, X.-B., {et~al.} 2014, \apjl, 795, L29,
  \dodoi{10.1088/2041-8205/795/2/L29}

\bibitem[{{Zhang} {et~al.}(2013){Zhang}, {Zhang}, \&
  {Liang}}]{2013ApJ...767....8Z}
{Zhang}, J., {Zhang}, S.-N., \& {Liang}, E.-W. 2013, \apj, 767, 8,
  \dodoi{10.1088/0004-637X/767/1/8}

\end{thebibliography}

\end{document}